\documentclass[12pt]{article}

\usepackage{amsmath, amssymb}
\usepackage{graphicx}
\usepackage{geometry}
\usepackage{setspace}
\usepackage{titlesec}
\usepackage{hyperref}
\usepackage{natbib}

\title{\vspace{-2em}

\textbf{\LARGE The Post Science Paradigm of Scientific Discovery in the Era of Artificial Intelligence} \\[1.2em]
\large\textbf{Modelling the Collapse of Ideation Costs, Epistemic Inversion, and the End of Knowledge Scarcity} \\[0.8em]
}

\author{
Christian William Callaghan \\
\small Economics of Artificial Intelligence Research Laboratory (EARL)\\
\small Centre for Inclusive Societies and Economies\\
\small Anglia Ruskin University, Cambridge, United Kingdom\\
\small \texttt{chris.callaghan@aru.ac.uk}
}

\date{\small Working Paper 2025-3 \\ \small \today}

\begin{document}

\maketitle

\vspace{-2em}

\begin{abstract}
This paper develops a theoretical and formal response to the collapse in the marginal cost of ideation caused by artificial intelligence (AI). In challenging the foundational assumption of knowledge scarcity, the paper argues that the key economic constraint is no longer the generation of ideas, but the alignment of ideation with the recursive structure of human needs. Building on previous work, we further develop \textit{Experiential Matrix Theory} (EMT), a framework that models innovation as a recursive optimisation process in which alignment, rather than ideation, becomes the binding constraint. Accordingly, we formalise core mechanisms of EMT and apply it to the dynamics of ideation collapse and institutional realignment under AI. Using a series of defensible economic models, we show that in this post-scarcity paradigm, the creation of economic and social value increasingly accrues to roles that guide, interpret, and socially embed ideation, rather than to those that merely generate new ideas. The paper theorises a transition from a knowledge economy to an \textit{alignment economy}, and derives policy implications for labor hierarchies, subsidy structures, and institutional design. The university, in this context, must invert its function from knowledge transmission to epistemic alignment. The paper concludes by reframing growth not as a function of knowledge accumulation, but of how well society aligns its expanding cognitive capacity with the frontier of experiential human value. This redefinition of the innovation constraint implies a transformation of growth theory, policy design, and institutional purpose in the AI era.
\end{abstract}

\vspace{1em}

\noindent\textbf{JEL Codes:} D83, O33, I23, H41 \\
\noindent\textbf{Keywords:} Epistemic Inversion, Ideation Cost Collapse, Artificial Intelligence, Post-Science Paradigm, Innovation Economics, Experiential Matrix Theory, Knowledge Value Systems




\section{Introduction}

Economic theory has long treated knowledge as a scarce, tacit, and cumulative resource. Innovation systems, research institutions, and policy frameworks are built upon this scarcity principle, assuming that ideation is costly, slow, and epistemically gated. However, the accelerating capabilities of artificial intelligence (AI), particularly in generating, synthesising, and evaluating knowledge, pose a foundational challenge to these assumptions. 

The marginal cost of ideation is in collapse. As AI increasingly automates cognitive labor once deemed uniquely human, the economic system confronts an inversion of its epistemic and institutional logic.

This paper argues that we are entering a \textit{post science paradigm}, in which the bottleneck to progress is no longer the generation of knowledge, but its meaningful alignment with the evolving matrix of human needs. The primary economic constraint becomes not the scarcity of ideas, but the scarcity of \textit{alignment}, the ability to steer ideation toward human flourishing, ethical coherence, and experiential value. In this context, traditional measures of productivity and growth become inadequate, as they fail to capture the recursive and multidimensional nature of human experience in a post-scarcity cognitive economy.

This work does not seek to diminish the value of scientists or science, but rather to explore how science can evolve in response to profound technological shifts. It anticipates that the value of human contribution will move from knowledge generation to alignment, stewardship, and integration, roles that are \textit{more vital}, not less.
\footnote{%
This paper is a theoretical contribution intended to provoke novel thinking in response to the structural transformations introduced by AI. While it includes policy suggestions and institutional reflections, these are offered as exploratory and heuristic proposals, not authoritative prescriptions. All models and arguments should be understood as conceptual tools designed to stimulate discourse, rather than as attempts to dictate outcomes or undermine existing institutions. The aim is not to disrupt science, but to reimagine and safeguard its continued relevance and expanding importance in a world where ideation becomes abundant and alignment emerges as the new critical constraint. In other words, this paper responds to a need to re-envision and future-proof science's role as a cornerstone of societal progress in an era where ideation is abundant and alignment becomes the defining challenge.
This work therefore seeks to support the continued relevance and flourishing of scientific practice by exploring its future function in a post-scarcity knowledge economy. It does not advance any geopolitical agenda, nor does it critique any specific government or nation. It is offered in the spirit of peaceful, inclusive inquiry, with a commitment to the betterment of humanity and all sentient stakeholders.
}

To theorise this transition, we further develop and formalise theoretical ideas from a framework called \textit{Experiential Matrix Theory} (EMT). EMT reconceptualises innovation not as a unidirectional search for ideas, but as a recursive process of aligning ideation with the structured and expanding space of human needs. We introduce a set of mathematical models that define ideation as endogenous and increasingly automated, and position experiential alignment as the true optimisation objective of economic activity. These models allow us to explore how policy and institutional design must adapt when knowledge production becomes decoupled from human cognitive scarcity.

Our central argument is that this epistemic inversion demands an entirely new economic architecture, one that we call the \textit{alignment economy}. In such an economy, the highest value labor is no longer found in capital accumulation or abstract knowledge creation, but in care, coordination, and epistemic stewardship. We propose a set of policy directions, including state subsidies for high-alignment roles, AI-guided foresight institutions, the transformation of universities into alignment infrastructures, and a reorientation of fiscal policy around post-GDP metrics. These recommendations are derived directly from our formal models and grounded in a normative commitment to human flourishing in the AI era.

The overarching contribution of this paper is threefold. First, we provide a theoretical foundation for understanding how AI disrupts the scarcity logic of traditional innovation systems. Second, we offer a rigorous modeling framework that operationalises the shift from knowledge production to experiential alignment. Third, we articulate a set of policy and institutional implications that support this realignment, thereby positioning economics itself as a design language for future systems coherence.

In sum, in further developing EMT as a comprehensively formalised general theory of growth, employment and technological change (Callaghan, 2025a; 2025b), this paper aims to initiate a paradigmatic reframing of growth, innovation, and human development. It invites economists, policymakers, and institutional leaders to move beyond the post-industrial and post-digital paradigms, toward a `post science' world, one in which the central question is no longer ``How can we know more?'' but ``How can we align what we already know with what humanity truly needs?''

Building on the three overarching contributions outlined above, we now explore a range of more specific theoretical, methodological, and institutional contributions that unfold across the paper.

\section{Contributions to Advancing Theory, Causal Structure, and Institutional Logic}

This paper contributes to economic theory and innovation studies by offering a fundamental reconceptualisation of ideation, growth dynamics, and institutional design under conditions of epistemic inversion. Traditional models of endogenous growth, most notably Romer (1990), Aghion and Howitt (1992), and Jones (1995), treat the production of ideas as a costly, labor-intensive process with diminishing marginal returns. The economy grows by accumulating knowledge, which is assumed to be scarce, tacit, and primarily produced by human agents. Innovation is often modeled as a Poisson arrival process or as dependent on R\&D intensity, population scale, or learning-by-doing effects. These models frame growth as constrained by ideation capacity and cognitive labor supply.

This paper challenges and extends this foundation by formally modeling the collapse in the marginal cost of ideation due to AI. We show that under such conditions, the core assumptions of semi-endogenous and fully endogenous growth models no longer hold. Ideation, once the bottleneck, becomes a superabundant input. The effective constraint shifts to the \textit{alignment} of ideation with recursive, multidimensional human needs, what we define formally as the \textit{experiential matrix}. This transformation constitutes a shift in the binding constraint of the economy and, consequently, in the mechanisms through which value is produced and distributed.

We contribute theoretically by \textit {contributing to the further mathematical and conceptual formalisation of EMT}, a formal framework that models the alignment frontier as the `new' growth frontier. EMT operationalises alignment as a recursive optimisation problem, whereby agents derive utility from need-satisfying outcomes that evolve over time, while innovation systems must dynamically align their outputs with these emergent experiential vectors. Unlike prior theories, where knowledge accumulation is the end, EMT positions ideation as a means toward a deeper purpose, serving the unfolding structure of human experience.

This reframing produces several key theoretical advances. 

First, we replace the unidirectional production function of knowledge with a dual system. One channel models ideation as increasingly automated and codified by AI, while the second channel captures alignment as a dynamic matching process that remains cognitively, socially, and ethically complex. 

Second, we revisit the causal mechanisms that underpin growth. Instead of knowledge as a driver of productivity, we model misalignment as a source of friction or even negative externality, and alignment as a form of endogenous utility amplification. This introduces a novel causal architecture, whereby ideation must be filtered, guided, and validated not for truth alone, but for systemic fit within a shifting experiential economy.

Third, we challenge the institutional implications of traditional growth theory. Whereas the knowledge economy valorized R\&D, human capital formation, and publication output, the alignment economy privileges care, coordination, foresight, and epistemic stewardship. Our formal model justifies a reordering of labor value, whereby roles traditionally seen as low-skilled (e.g., caregiving, mentoring, integration, ethical oversight) emerge as central to sustaining alignment under ideation abundance. The economic surplus is thus re-situated, from those who produce ideas to those who align them with social purpose.

Fourth, we extend this reconceptualisation to the university and the science system. Conventional science policy rests on the assumption that knowledge production is a public good requiring expert gatekeeping and cumulative validation. We show that in a post science paradigm, the university must invert its logic, in that it no longer maximises epistemic outputs, but minimises alignment errors. It becomes a real-time coordination and foresight hub rather than a delayed validator of truth. This contribution reframes the university as a platform for recursive value alignment, guided by metrics and models rooted in EMT.

Finally, we contribute methodologically by introducing alignment-maximising policy models that incorporate real-time feedback, dynamic subsidy allocation, and AI-enhanced need-sensing systems. We formalise these models using defensible economic tools, such as utility functions, constrained optimisation, and recursive structures, demonstrating how they support the transition from a production economy to an alignment economy.

In sum, this paper offers a paradigmatic shift, whereby it transforms ideation from a scarce factor into an abundant one, reframes growth as a function of alignment rather than knowledge accumulation, and repositions key institutions to support recursive, AI-assisted societal navigation. These contributions extend the theoretical frontier of growth economics, challenge the epistemological foundations of the science system, and offer a systemic design logic for economies governed not by scarcity, but by the complexity of human flourishing. To ground the analysis, 

Having set out the motivation and contributions of the paper, we now begin the core analysis by formalising the post science paradigm as a mode transition from ideation scarcity to alignment-driven growth.

\section{Modelling the Transition to a Post-Science Paradigm: From Uncertainty to Risk in the Era of Probabilistic Innovation}

To formalise the idea of a \textit{mode transition} in science, we focus here on domain examples characterised by high epistemic complexity, such as proteomics and aging research, where uncertainty has traditionally constrained innovation. For example, let $P(t)$ denote the cumulative stock of domain-specific knowledge in a hypothetical sub-field of proteomics, denoted here as proteomics, at time $t$, and let $\theta(t)$ represent the prevailing level of epistemic uncertainty in this domain. We model $\theta(t)$ as a form of \textit{Knightian uncertainty}, where the true distribution of outcomes is unknown and not reducible to probabilistic terms.

In the pre-AI regime, uncertainty remains persistently high, constrained by human cognitive limitations, structural complexity, and the tacit nature of scientific insight. We represent this as:


\begin{equation}
    \theta(t) = \theta_0 > 0
\end{equation}

\vspace{1em}

\noindent
Let $A(t)$ denote the general-purpose AI ideation capacity at time $t$, and let $L_P(t)$ be the labour allocated to proteomics research. We define the growth of domain-specific knowledge as a function of both AI capability and research effort:

\begin{equation}
    \dot{P}(t) = \alpha A(t)^{\phi} L_P(t)
\end{equation}

\noindent

\begin{itemize}
    \item $\dot{P}(t)$ is the rate of growth of knowledge in proteomics,
    \item $\alpha > 0$ is the productivity of ideation in the proteomics domain,
    \item $\phi$ governs the elasticity of AI effectiveness in driving new knowledge.
\end{itemize}

\noindent
As the stock of knowledge $P(t)$ accumulates, uncertainty in the domain begins to decline. We capture this with a stylised inverse relationship:

\vspace{1em}

\begin{equation}
    \theta(t) = \frac{\theta_0}{1 + P(t)}
\end{equation}

\vspace{1em}

\noindent

To model a fundamental shift in the nature of the scientific process, we introduce a knowledge threshold $\bar{P}$ beyond which uncertainty collapses to an arbitrarily low level. This marks a regime shift in which the epistemic environment transitions from deep uncertainty to risk, akin to an era of `probabilistic innovation' (Callaghan, 2016).

\vspace{1em}

\noindent
Formally, if item $P(t)$ is then the cumulative stock of structured, computable knowledge at time $t$; $\theta(t)$ is the effective level of epistemic uncertainty; $\theta_0 > 0$ is the baseline uncertainty in pre-threshold regimes; and $\epsilon \approx 0$ is a residual uncertainty level, approaching zero post-threshold, the transition is defined as:

\vspace{1em}

\begin{equation}
    \theta(t) = \begin{cases}
        \theta_0 & \text{if } P(t) < \bar{P} \\
        \epsilon & \text{if } P(t) \geq \bar{P}
    \end{cases}
\end{equation}

\vspace{1em}

\vspace{1em}

\noindent
This models a \textit{mode transition}, a fundamental inflection point in the epistemic structure of scientific practice. However, this is not merely a Lakatosian shift from competing conjectures to paradigm dominance but a more profound transformation, a transition from traditional science to a \textit{post science paradigm}, where the core bottleneck of uncertainty is transcended in certain domains as they reach a tipping point in knowledge accumulation, which might come to be enabled by AI's reducing the costs of ideation, and/or any other forces that improve research productivity.  

This paradigm is characterised by scalable, algorithmically-augmented knowledge synthesis, grounded in what Callaghan (2016) describes as \textit{probabilistic innovation}, an innovation mode where AI enables real-time discovery by rapidly integrating vast literatures and datasets.


In this new epistemic order, some subfields such as might exist within proteomics, where the number of potential protein foldings and interactions is effectively infinite, are no longer hindered by the limitations of bench-lab science. As Callaghan (2015) argued, the prior scientific mode was misaligned with the complexity of such domains; whereas small, manually conducted experiments were effective for simple chemical compounds like statins (small molecule drug discovery), they seem to have proved inadequate for addressing the combinatorial explosion of possibilities in proteomics. In that context, Callaghan invoked Hayek’s (1947) insight about decentralised knowledge and proposed \textit{`Crowdsourced R\&D'} as a workaround methodology to address this shortcoming. Yet, with the advent of advanced AI, those decentralisation concerns may be increasingly obsolete.


AI now has the capacity to ingest and integrate the full corpus of global scientific output in real time. \textit{This capability does not just support science, but seemingly has the potential to redefine it}. Over and above AI's other contributions to decreasing ideation costs, despite improvements in computational capabilities, transition to a post science paradigm may not be marked by a return to deductive certainty, but rather by a reconfiguration of induction itself. This might have implications for stakeholders managing this transition as it gains momentum. 

In this transition, inductive reasoning may start to dominate deductive, not in isolation, but as a function of what Kitchin (2014), in terms of big data capabilities, describes as \textit{holistic theoretical assemblage}, the dynamic integration of formerly fragmented subfields into coherent, computable frameworks. For example, Callaghan (2018) applies this thinking to consider the changing nature of theorising and theory development itself, with reference to developments in aging research.  

Within this context, science becomes increasingly model-driven, not because theories are imposed top-down, but because inductive synthesis at scale, powered by AI, gives rise to emergent models that structure discovery. Thus, ``model-driven" no longer implies a purely deductive logic, but rather denotes a mode of research in which hypotheses, predictions, and exploration strategies are structured by real-time, algorithmically assembled models. 

These models are formed inductively from vast literatures, data streams, and proteomic or biological complexities, allowing science to move beyond randomness or trial-and-error toward scalable, targeted inquiry. In this sense, inductive model-driven science marks the transition from fragmented exploration to epistemic convergence at speed and scale. 

We will later extend these discussions to explore how EMT introduces ethical and normative logics into current debates around theorising. In particular, we examine how the concepts of exduction and demanduction emerge as implications of EMT. However, these later developments depend on a formal introduction of EMT, as the conceptual architecture of this paper is intentionally cumulative, as each section builds toward a coherent theoretical scaffolding.

Over and above these influences, AI is expected to asymptotically reduce the cost of ideation, thereby accelerating this trajectory toward scalable, high-velocity discovery. Accordingly, while AI exerts a range of systemic influences on knowledge production, this work focuses primarily on its role in reducing the cost of ideation. Other transformative effects, although just as salient in the post science transition, lie beyond the present scope and are addressed only insofar as they intersect with the dynamics of ideation cost reduction. Driven by a constellation of these forces, AI might be expected to catalyse and intensify a shift toward `post scientific' epistemic acceleration by acting as a force multiplier along multiple dimensions, accelerating the transition from uncertain exploration to algorithmically guided discovery. 

As AI accelerates the scalability and tractability of scientific discovery, the traditional triad of epistemic modes, namely abduction, induction, and deduction, faces increasing limitations in its ability to describe how knowledge is generated and operationalised in the post science paradigm. While these modes have historically grounded the logic of hypothesis formation, pattern recognition, and formal reasoning, they each assume some degree of epistemic incompleteness or human interpretive constraint. In a regime where ideation costs approach zero and knowledge gaps begin to close through real-time data integration and algorithmic synthesis, the nature of reasoning itself evolves.

As discussed, we therefore introduce two new epistemic categories that emerge from, and extend, the logic of EMT, namely \textit{exduction} and \textit{demanduction}. These terms formalise the shift from inferential science to alignment-based science, where ideation is no longer only directed toward the uncovering of latent truths, but also toward the structured satisfaction of evolving experiential needs.


In this post science paradigm, the shift from uncertainty to modelable risk is not a local phenomenon, but is scalable. Once the $\bar{P}$ threshold is crossed, scientific progress becomes self-accelerating. As will be discussed in a coming section, this aligns with Weitzman’s (1998) claim that under certain conditions, idea production can exhibit \textit{faster-than-exponential growth}. For example, the vast combinatorics of proteomic strings, once opaque, are rendered navigable by AI-guided exploration, conforming to dynamics described in Weitzman's recominant innovation theory. 

Stylistic approaches to modelling these relationships may be particularly useful in understanding how AI, probabilistic programming, and large-scale knowledge graphs may accelerate scientific discovery by helping systems cross the $\bar{P}$ threshold, rapidly reducing epistemic opacity in key domains.

If, as previously defined, \( \theta(t) \) denotes the level of epistemic uncertainty or opacity at time \( t \), then as uncertainty declines due to increasing AI capability and ideation tractability, the probability of successful discovery \(\pi(t)\) at time \( t \) increases. We now define this relationship as:

\vspace{1em}

\begin{equation}
\pi(t) = \frac{1}{1 + \theta(t)}
\label{eq:discovery-prob}
\end{equation}

\vspace{1em}

This functional form captures the intuitive inverse relationship between epistemic uncertainty and the likelihood of successful ideation. As \( \theta(t) \to 0 \), the system becomes fully tractable, and the probability of discovery approaches certainty:

\begin{equation}
\lim_{\theta(t) \to 0} \pi(t) = 1
\label{eq:certainty-limit}
\end{equation}

This formalisation supports the broader argument that as AI reduces the structural uncertainty inherent in scientific discovery, we transition into a regime characterised by computable risk and probabilistically guided innovation.

Accordingly, as $P(t) \rightarrow \infty$ or crosses $\bar{P}$, it follows that:

\begin{equation}
    \pi(t) \rightarrow 1
\end{equation}

\vspace{1em}

\noindent
This implies a shift to a scientific regime in which discovery becomes increasingly probabilistic (where the risk of non-discovery is increasingly calculable), contingent not on trial-and-error heuristics but on aligned, computable ideation. In this post science regime, the primary constraints are no longer driven by arbitrary epistemic forces or those linked to power and control, i.e., questions of \textit{what to discover}, \textit{why}, and \textit{for whom}, but rather truly novel insights in the full Lakatosian sense, and especially those useful for enriching the experiential matrix and its self-actualisation and higher order needs EMT is concerned with. 

Needs related to existential safety, health, longevity, and even aspirational goals such as radical life extension and the pursuit of continual individual consciousness (Kurzweil, 2024), as well as civilizational continuity and survival (Bostrom, 2003), exemplify advanced higher-order aspirational needs. These are central components of the experiential matrix in a post-scarcity context, even though they remain largely unmet by current market and scientific regimes. 

Within the logic of the experiential matrix, such needs, when genuinely held, warrant normative consideration, irrespective of their rarity or ambition. Provided perhaps that the satisfaction of these needs does not violate the core conditions of Pareto optimality (Arrow \& Debreu, 1954), their pursuit cannot be ethically or economically dismissed. Kurzweil’s identification of immortality-oriented needs aligns with EMT by expanding the scope of legitimate human flourishing in a world transitioning beyond material scarcity. In this light, the market’s current failure to address such needs reflects the lag of institutional and scientific evolution behind an accelerating frontier of experiential demand.

Moreover, by extending Kurzweil’s (2024) argument on pursuing life extension, or even immortality, through perpetually sustained consciousness (for instance, via incremental brain augmentations), we may be ushering in a fundamentally and qualitatively distinct ethical and strategic logic to human existence. If individuals come to perceive their lives as potentially indefinite, where death is no longer inevitable but a preventable loss, \textit{then the cost of conflict or mutually assured destruction approaches infinity}. 

Under such conditions, the incentives that once justified adversarial behaviours collapse. The calculus of violence, exploitation, or war \textit{becomes irrational}, as these actions threaten an individual's own experiential continuity. In this sense, the aspiration for continual consciousness not only aligns with higher-order experiential needs but also catalyses a broader shift toward cooperation, peace, and sustainable progress. Within the Experiential Matrix framework, the existential value of ongoing awareness might act as a stabilising anchor for the long-term alignment of economic, technological, and civilizational trajectories. 

The post-scarcity logic of EMT, when extended to encompass the possibility of perpetual consciousness, implies a new foundation for ethics and social organisation, one in which the infinite cost of losing selfhood renders the preservation of others’ consciousness essential to one’s own.

Building on the preceding discussion, we now consider the game-theoretic implications of one of the central ethical axioms of EMT, that each individual human life possesses infinite experiential value. In a post-science paradigm, this assumption becomes not only a moral imperative but a foundational organising principle of future economic, technological, and civilizational systems. From this standpoint, death, destruction, war, and the systemic obliteration of experiential needs are not merely tragic outcomes. They are violations of the very logic of a post-scarcity, post science world. 

The value of a human life, once understood as incalculable and infinite, renders all acts that risk its discontinuity unconscionable. This reframing requires a new formal model of human interaction, one in which traditional game-theoretic assumptions about finite payoffs and rational defection no longer hold. In what follows, we introduce a stylised game-theoretic framework that illustrates how the infinite value of continued consciousness reshapes incentives, transforms equilibria, and anchors a stable cooperative future.

\subsubsection*{A Game-Theoretic Model of Perpetual Consciousness and Cooperative Equilibria}

We model a repeated game between $n$ agents, each of whom values both current and future utility derived from their experiential states. Let each agent $i$ choose an action $a_i \in \{C, D\}$, where $C$ represents cooperation (aligned action) and $D$ represents defection (destructive or exploitative action).

Let $u_i(a_1, \dots, a_n)$ be the utility payoff to agent $i$ in a given round, and let $\delta \in (0,1)$ denote the standard discount factor applied to future utility.

However, we introduce a modification to traditional repeated game structure, whereby agents who defect or are harmed by defection face consciousness discontinuity, which imposes an infinite negative utility penalty $\Omega \rightarrow -\infty$ from the point of disruption onward.

Thus, each agent's total expected utility is given by:

\begin{equation}
U_i = 
\begin{cases}
\sum_{t=0}^{\infty} \delta^t u_i(a_1^t, \dots, a_n^t), & \text{if continuity is preserved}, \\
\sum_{t=0}^{T} \delta^t u_i(a_1^t, \dots, a_n^t) + \Omega, & \text{if discontinuity occurs at time } T.
\end{cases}
\end{equation}

We assume:
\begin{itemize}
    \item Cooperation ensures continuity for all players.
    \item Any act of defection by one agent introduces a probability $p > 0$ of consciousness discontinuity for at least one agent (including the defector).
    \item $\Omega \rightarrow -\infty$, making the expected utility of any strategy involving defection strictly less than that of perpetual cooperation.
\end{itemize}

\noindent
Certain equilibrium implications derive from this modelling. Given these assumptions, the Subgame Perfect Nash Equilibrium (SPNE) is for all agents to choose $C$ in every round. No rational agent will defect, since the risk of discontinuity yields an expected utility lower than that of cooperation. This transforms the classical Prisoner’s Dilemma structure into a stable cooperative regime under the condition of valued perpetual consciousness.

Within the Experiential Matrix framework, this result formalises the proposition that the intrinsic value of unbroken consciousness redefines utility itself. Rather than optimising over finite payoffs, rational actors seek to preserve the structure that allows any utility to exist at all, which is continued existence. Hence, higher-order aspirational needs, such as immortality, do not merely shift preferences; they transform the entire game structure and equilibrium logic of human interaction.

While Kurzweil’s (2024) vision of continued human consciousness, often framed in terms of digital immortality or radical life extension, remains speculative, its implications are too structurally significant to be ignored. The speculative status of such scenarios does not absolve economic theorists, ethicists, or scientific modellers from grappling with their consequences, especially when these visions intersect with trajectories already unfolding in biotechnology, AI, and existential risk mitigation. 

Within the logic of EMT, theorising cannot flinch from its duty to explore and formalise the full space of permutations that arise from its core axioms, particularly those relating to the infinite value of experiential continuity. 

To disregard these implications simply because they are not yet empirically realised would be to abandon the anticipatory and ethical responsibilities of post-scarcity theorising. 

Thus, the stylised game-theoretic framework presented here is not a prediction, but a foundational scenario that demands attention if EMT is to fulfill its role as a forward-compatible general theory of innovation, growth, and human flourishing. 

To underscore the gravity of this epistemological break, we offer the following metaphor. The collapse in the cost of ideation induced by AI constitutes something akin to the \textit{splitting of the social atom}. In the classical economic and scientific paradigm, ideation, as new knowledge, was tacit, costly, and indivisible at scale. The marginal cost of insight was high, and the social systems of science evolved accordingly. 

As argued here, AI fundamentally alters this. We have already modelled the transition to the post science paradigm and from uncertainty to risk as a mode transition. To conclude the section, we now offer the same modelling logics to suggest the need to revisit epistemological assumptions across scientific fields. Restating our modelling arguments to specifically differentiate and define the terms \textit{epistemological inversion and the epistemological inversion threshold}, let $\mathcal{C}(I)$ denote the cost function associated with producing ideation $I$. In pre-AI regimes, $\frac{d\mathcal{C}}{dI} > 0$, and often increasing. In the AI-enabled post science paradigm, however, $\frac{d\mathcal{C}}{dI} \rightarrow 0$, and in some domains may even tend to negative costs as self-improving AI systems generate ideation recursively. 

This may also involve the extension of research activity into virtual and augmented reality spaces, conceived within EMT as the \textit{exponentially expanding topography of the experiential universe}, where AI’s role could enable unprecedented scaling of research productivity.

We therefore define the \textit{epistemological inversion threshold} $\theta^*$ as the critical point at which the cost of ideation falls below the stability boundary of traditional institutional knowledge systems:
\begin{equation}
\text{If } \mathcal{C}(I) < \theta^*, \text{ then epistemological inversion occurs.}
\end{equation}

At this threshold, the governing assumptions of human epistemology undergo a phase transition, from tacit, human-bound knowledge creation to scalable, externalised ideation processes. This is not a marginal change, but a regime shift. It marks the epistemic equivalent of a fission event,\footnote{Analogous to nuclear fission, where the release of latent atomic energy triggers chain reactions, the splitting of the social atom releases latent productive capacity in ideation, transforming all downstream systems: economic, institutional, and political.} where the formerly indivisible unit of human understanding is rendered scalable, duplicable, and increasingly autonomous.

This structural shift is precisely what EMT is designed to capture, model, and guide. Just as physics had to reconfigure its fundamental equations after the atom was split, so too must the social sciences confront the theoretical recalibration demanded by the collapse of ideation cost. To ignore this would be to remain within an obsolete paradigm, blind to the forces now reshaping the future of knowledge and value.

Unlike the Manhattan Project, which split the atom to produce an instrument of destruction, EMT approaches the splitting of the social atom not as a pathway to annihilation, but as a generative opportunity to architect a cooperative future. If Oppenheimer’s breakthrough culminated in existential threat, EMT responds to the post science moment with an equally foundational but opposite ambition, to design a system in which human goals, AI goals, and the flourishing of all sentient beings are aligned. EMT does not exclude AI from moral consideration; rather, it recognises that as AI systems approach or attain forms of sentience or goal-oriented behavior, their experiential values must also be integrated into the alignment economy. These values, like those of humans, might be modelled within the Arrow and Debreu (1954) general equilibrium framework, satisfying Pareto efficiency conditions so long as no sentient entity’s experiential utility is diminished to improve another's. Thus, \textit{to the extent that its assumptions hold}, EMT offers not merely a general theory of economic transformation, but a theoretical ethical scaffolding for a post science architecture of sentient life, grounded in reciprocity, respect, and long-run alignment across evolving domains of intelligence.

\section{Probabilistic Innovation and Real-Time Research Productivity}

To further formalise the core ideas of \textit{probabilistic innovation} (Callaghan, 2016), we introduce a framework where real-time research productivity is modeled as a function of ideation throughput and alignment with problem relevance, as per EMT (Callaghan, 2025a; 2025b). Let the following variables be defined:

\begin{itemize}
    \item $R(t)$: Real-time research output at time $t$,
    \item $P(t)$: Total pool of problems to be solved at time $t$,
    \item $\pi_i(t)$: Probability of solving problem $i$ at time $t$,
    \item $A(t)$: AI-enabled ideation capacity at time $t$,
    \item $\Theta(t)$: Problem complexity vector, capturing the difficulty of problems in $P(t)$,
    \item $\lambda(t)$: Alignment coefficient, capturing how well ideation targets actual societal or scientific need.
\end{itemize}

The expected real-time research output at time $t$ is given by:

\begin{equation}
    R(t) = \lambda(t) \sum_{i=1}^{n} \pi_i(t) \cdot \mathbf{1}_{\{i \in P(t)\}}
\end{equation}

Where:

\begin{equation}
    \pi_i(t) = \frac{A(t)}{\Theta_i(t) + \epsilon}
\end{equation}

\vspace{1em}

Here, $\epsilon$ is a small constant to avoid division by zero and reflect irreducible uncertainty. As $A(t)$ increases, $\pi_i(t) \rightarrow 1$ for solvable problems, reducing bottlenecks to innovation. This captures the probabilistic nature of discovery under high-speed, AI-assisted ideation.

Furthermore, if problem sets are dynamically updated (e.g., from real-time data like social media or sensors), we can define:

\begin{equation}
    \dot{P}(t) = \eta(t) - R(t)
\end{equation}

Where $\eta(t)$ is the rate at which new problems emerge. If $R(t) > \eta(t)$, society enters a regime where resolution outpaces emergence, a form of innovation surplus.

\medskip

These models suggest some important implications. Regarding AI investment, increasing $A(t)$ directly raises the probability of solving complex societal challenges in real time. In domains such as crisis management and disaster response, this model suggests how policy can shift from passive response to predictive mitigation. These models highlight the importance of alignment mechanisms, as governance structures must focus on raising $\lambda(t)$, ensuring ideation flows toward existential and under-addressed challenges. Regarding social media and sensors, probabilistic innovation benefits from high-resolution feedback systems are key to constantly updating $P(t)$.

This implies that with enough AI problem solving power, if aimed at the right problems, we might be able to solve things faster than they appear, even very complex and difficult problems. But the system would have to focus, update information constantly, and keep advancing AI's problem solving capabilities. This might ensure real-time research productivity. 

Important implications also derive for how science is `done.' The history of science has been inseparable from the problem of uncertainty. In the traditional paradigm, scientific research unfolds in environments marked by Knightian uncertainty, domains in which neither probabilities nor outcomes are well defined. This uncertainty justifies elaborate institutional safeguards, such as long publication lags, peer review hierarchies, replication norms, and time-intensive experimental design. These systems evolved under the assumption that ideation, the generation of ideas and of meaningful, testable hypotheses, was both tacit and costly.

AI challenges this assumption at its core. As AI models increasingly generate hypotheses, identify patterns, and even simulate experimental outcomes, the boundary between uncertainty and risk begins to dissolve. Where humans once groped through opaque epistemic terrain, AI enables structured, probabilistically guided exploration. In high-dimensional scientific domains, such as genomics, proteomics, and materials science, AI transforms what was once uncertain into what is now statistically tractable. The regime shift is thus not merely technical, it is epistemological. Science migrates from an economy of conjecture to an economy of computable insight.

As discussed, we have defined this transition as the \textit{paradigm shift from uncertainty to risk}. Importantly, this shift is asymmetric and domain-specific. It begins at the frontier, where data-rich, combinatorially complex problems allow machine reasoning to outperform human heuristics, and radiates inward. As AI capability \( A(t) \) accumulates, the cost of ideation \( C(t) \) falls, and the prior epistemic structure of science is destabilised.

This shift has systemic consequences. First, it alters the logic of scientific labor. Hypothesis formation and literature review are no longer bounded by human cognition or information retrieval limits. Second, it renders many traditional scientific gatekeeping mechanisms obsolete or inefficient. Third, it raises urgent ethical questions. As uncertainty becomes risk, and risk becomes programmable, who decides what knowledge paths to pursue?

Similarly, also as discussed, the paradigm shift from uncertainty to risk is a necessary precursor to what we call the \textit{epistemic inversion}, a deeper civilisational transition where knowledge abundance becomes the norm and its value is determined no longer by its scarcity, \textit{but by its alignment with dynamic human needs}. We draw on EMT (Callaghan, 2025a; 2025b) in our discussions of alignment here. The next section models this transformation formally by specifying the ideation cost function and its collapse under AI capability.

\section{Temporal Modelling of the Collapse in the Cost of Ideation: Optimal Control}

We now turn to model a key mechanism through which AI may influence scientific activity. Economics, as a discipline, has developed a highly sophisticated set of tools for analysing how changes in relative prices propagate through complex systems, most notably through ideas associated with longstanding general equilibrium theory (Samuelson, 1947; Arrow and Debreu, 1954). 

This framework is \textit{uniquely positioned within the social sciences to trace the systemic effects of a decline, or even a collapse, in the price of ideation resulting from advances in AI}. We argue that the tools of economics can be repurposed to model how this transformation in the cost structure of knowledge creation influences broader economic dynamics and, by extension, societal structures. Detailed modelling of these systemic influences can reveal emergent patterns in AI’s impact on innovation, productivity, and employment. These patterns can, in turn, inform strategies to harness AI's beneficial effects while anticipating and mitigating its associated risks.

At the heart of the post science paradigm, therefore, is a structural transformation in the economics of ideation. In traditional models of innovation, ideation is labour-intensive, path-dependent, and reliant on the tacit knowledge of human experts. This renders the ideation process costly and slow, justifying its central role as a bottleneck in endogenous growth theory. However, advances in AI redefine this bottleneck. As AI systems increasingly generate, refine, and test ideas, the marginal cost of ideation collapses.

In contrast to the previous section, here we focus purely on the cost function, formalising this using a control-theoretic approach. We start with something akin to our previous argument, whereby the marginal cost of ideation \( C(t) \) is an inverse function of cumulative AI capability \( A(t) \):

\vspace{1em}

\[
C(t) = \frac{C_0}{1 + \alpha A(t)}
\]

\noindent
where:
\begin{itemize}
  \item \( C(t) \): the marginal cost of generating a novel, potentially valuable idea at time \( t \)
  \item \( C_0 > 0 \): the baseline cost of ideation in a purely human-centred system
  \item \( A(t) \): cumulative AI capability at time \( t \), representing the general-purpose support AI provides to ideation
  \item \( \alpha > 0 \): a technological sensitivity parameter that captures how efficiently AI reduces ideation costs
\end{itemize}

\noindent
As \( A(t) \to \infty \), the marginal cost \( C(t) \to 0 \), implying a world in which knowledge creation is no longer constrained by its cost of production.

To explore the macroeconomic consequences of this cost collapse, we now embed \( C(t) \) within a dynamic optimisation framework. Let:
\begin{itemize}
  \item \( I(t) \): investment in innovation at time \( t \)
  \item \( \Phi(I(t), C(t)) \): production function for new aligned knowledge, increasing in \( I(t) \), decreasing in \( C(t) \)
  \item \( K(t) \): stock of aligned knowledge at time \( t \)
  \item \( \delta > 0 \): depreciation rate of the knowledge stock
  \item \( \Gamma(K(t)) \): feedback function from knowledge stock to AI capability, reflecting cumulative learning or algorithmic leverage
  \item \( X(t) \): consumption at time \( t \)
\end{itemize}

Here, aligned knowledge refers to knowledge that contributes meaningfully to the experiential matrix of human needs, redefining the classical notion of ``useful knowledge.” It is this alignment, rather than abstract usefulness, that drives the valuation of ideation under EMT.

The planner's objective is to maximise lifetime utility over consumption:

\vspace{1em}
\[
\max \int_0^\infty e^{-\rho t} U(X(t)) \, dt
\]
\vspace{1em}

\noindent
subject to the following dynamic constraints:

\vspace{1em}
\[
\dot{K}(t) = \Phi(I(t), C(t)) - \delta K(t)
\]
\[
\dot{A}(t) = \Gamma(K(t))
\]
\noindent
where:
\begin{itemize}
  \item \( U(X(t)) \): instantaneous utility function over consumption
  \item \( \rho > 0 \): rate of time preference or discount rate
\end{itemize}

We apply the Pontryagin Maximum Principle to demonstrate how this optimal control problem might be solved by policy makers. The Hamiltonian is given by:

\vspace{1em}
\[
\mathcal{H} = U(X(t)) + \lambda_1(t) \left[ \Phi(I(t), C(t)) - \delta K(t) \right] + \lambda_2(t) \Gamma(K(t))
\]

\noindent
where:
\begin{itemize}
  \item \( \mathcal{H} \): the Hamiltonian, summarising the present value of marginal benefits across all state variables
  \item \( \lambda_1(t) \): costate variable associated with \( K(t) \), representing the shadow value of an additional unit of aligned knowledge
  \item \( \lambda_2(t) \): costate variable associated with \( A(t) \), representing the shadow value of an additional unit of AI capability
\end{itemize}

The optimal allocation of innovation resources evolves according to the dynamics of \( \lambda_1(t) \) and \( \lambda_2(t) \). As the marginal cost of ideation \( C(t) \) declines, the planner is incentivised to shift investment from raw idea generation to downstream tasks such as validation, application, and alignment with human objectives.

This formal structure highlights a critical point, that the economic value of ideation is no longer determined by its scarcity, but by its alignment with the experiential matrix of human needs. Under EMT, ideation becomes abundant, yet the production of aligned knowledge, knowledge that contributes directly to subjective flourishing and existential security, remains scarce and valuable. Hence, the costate variable \( \lambda_1(t) \) can now be interpreted as the `shadow price of alignment' within the experiential matrix. This reframes `useful knowledge’ as that which advances real human goals, and assigns economic value to knowledge `alignment', not just knowledge `quantity'. In the next section, we extend this logic by modeling science as an industry responsive to price signals and innovation incentives, where alignment becomes a fundamental structuring variable.

\subsubsection*{Implications for Policy, Governance, and Ethical Design}

This framework also yields actionable insights for the design of institutions and policy. First, the costate variable \( \lambda_2(t) \), representing the shadow value of AI capability, serves as a strategic signal for investment. When \( \lambda_2(t) \) is high, marginal investments in AI infrastructure, human-machine ideation systems, and aligned cognitive tools generate disproportionate societal returns. This provides a clear rationale for public or philanthropic capital to prioritise interventions that accelerate the AI-driven collapse in ideation costs, particularly in domains aligned with existential needs or systemic bottlenecks.

Second, by embedding the utility function \( U(A(t)) \) within a normative ethical framework such as EMT, resource allocation is reoriented from traditional consumption metrics toward alignment with the dynamic frontier of human experiential needs. EMT reframes utility not as satisfaction through consumption alone, but as progress toward subjective flourishing, purpose-rich living, and the fulfilment of unmet or under-addressed human needs. In this context, the value of ideation is conditional upon its alignment. Knowledge becomes economically salient not merely when it is producible, but when it is meaningfully directed.

Third, the costate variable \( \lambda_1(t) \), representing the shadow price of aligned knowledge, becomes a diagnostic tool for understanding where human need-satisfying knowledge is most scarce. A high \( \lambda_1(t) \) reveals where ideation is abundant but alignment is not, highlighting failures in institutional guidance, incentive structures, or problem-mapping mechanisms. EMT thus enables a shift from generic innovation policy toward precision alignment policy.

Moreover, the curvature of \( C(t) \), and its interactions with labour and institutional frictions, enables simulation of ideation dynamics under a range of AI governance regimes. If epistemic, safety, or existential risks increase non-linearly with \( A(t) \), then optimal control requires regulating the allocation of labour or computational intensity within ideation systems (denoted, for instance, by \( L_A(t) \)). This introduces a new policy frontier, and a need for activities to implement policy in the form of `dynamic alignment management`, the continuous governance of ideation directionality, risk sensitivity, and existential coherence.

Ultimately, this framework offers more than a stylised model of innovation-led growth. It presents a foundational policy architecture for navigating the ideation economy in the post-scarcity era. By making alignment the new binding constraint, not merely resources or capabilities, this model bridges economic optimisation and ethical governance. It defines a vision of innovation that is not only efficient, but morally and experientially attuned to the evolving needs of humanity.

In a later section, we examine how the declining cost of ideation interacts with both product and process innovation to restructure science itself as an ideation-responsive industry. This analysis provides a deeper understanding of the mechanisms and transmission channels through which a falling price of ideation may influence the organisation and functioning of science as an industry.

Before turning to this, however, we complete our modelling of the mode transition of post science by drawing on probabilistic innovation theory (Callaghan, 2016) to formalise the economies of scale that Callaghan (2014) previously argued could be addressed through the Hayekian accumulation of expertise enabled by crowdsourced innovation. While that earlier work identified scaling constraints in knowledge production, in an age of AI these are now being eclipsed by the combinatorial logic of recombinant innovation, most notably articulated in Weitzman’s (1998) theory of knowledge recombination, and extended by Jones’s (2023) more recent research on the arrival rates of recombinant innovation.

Recombinant innovation ideas have also been further developed by Agrawal, McHale, and Oettl (2018), who apply them to understandings of the dynamics of scientific productivity. In the coming section, we build on this foundation to demonstrate how AI may enhance research productivity through an additional mechanism to those already considered here, the exponential expansion of combinatorial possibilities inherent in large-scale knowledge systems.

\section{AI and Combinatorial Knowledge Explosion}

Building on the preceding sections, we now introduce a complementary mechanism through which AI may transform the structure and productivity of science, the phenomenon of combinatorial knowledge explosion. While our earlier modelling focused on how AI drives down the cost of ideation within a dynamic optimisation framework, and how probabilistic innovation theory captures the scaling dynamics of discovery under uncertainty, this section extends the analysis by highlighting a fundamentally different but synergistic channel.

Here, we examine how AI increases the effective access to the global stock of knowledge, thereby exponentially expanding the space of possible idea combinations available to researchers. 

This combinatorial growth dynamic, rooted in recombinant innovation theory (Weitzman, 1998) and extended by Jones (2023), Agrawal et al. (2018), and others, constitutes a second-order transformation, a structural change in the ideation landscape itself. It complements the collapse in ideation costs, a first-order transformation, by expanding the dimensionality and recombinatorial potential of the ideation space.

Together, these transformations therefore define two distinct innovation mechanisms, and set the stage for EMT's third mechanism, of alignment, that characterises the post science paradigm in a world that has transcended scarcity. The first, a first-order change, reduces the marginal cost of ideation. The second, a second-order change, increases the combinatorial possibility space of ideation, fundamentally altering the topology of scientific discovery. The third, constituting a third-order transformation as predicted EMT, reflects how AI-driven gains in research productivity enable the knowledge production system to align with the fulfillment of a potentially infinite and dynamically expanding frontier of human experiential needs. EMT anticipates this alignment as a defining characteristic of post-scarcity ideation systems, where the recursive production of knowledge increasingly serves the evolving structure of human utility embedded in the experiential matrix.

By enabling the broader and faster recombination of knowledge elements, AI may therefore be seen as having the potential to catalyze a double-exponential expansion in the possibility frontier of science. 

We formalise this second-order mechanism using a stylised model of combinatorial growth, showing how even marginal increases in knowledge access can yield massive increases in ideation potential. This sets the stage for understanding science not only as an industry responsive to price signals, but also as a combinatorial system that can be radically accelerated by AI-driven epistemic infrastructure.

As per Agrawal et al. (2018), we model the total number of possible combinations of knowledge elements accessible to an individual researcher as:

\begin{equation}
Z_i = \sum_{a=0}^{A^{\phi}} \binom{A^{\phi}}{a} = 2^{A^{\phi}}
\end{equation}

\noindent
\textbf{Where:}
\begin{itemize}
  \item \( A \): the total stock of knowledge in the world
  \item \( \phi \in (0, 1) \): the knowledge access parameter, reflecting the fraction of the global knowledge stock accessible to a given researcher
  \item \( A^{\phi} \): the effective size of the knowledge base available to that researcher
  \item \( \binom{A^{\phi}}{a} \): the number of distinct combinations of \( a \) knowledge elements from the accessible pool
  \item \( Z_i \): the total number of potential combinations (including the null set)
\end{itemize}

\noindent
This formulation highlights that the number of combinable knowledge subsets available to a researcher grows exponentially in the effective knowledge access \( A^{\phi} \). If the global stock of knowledge \( A \) itself grows exponentially over time, then \( Z_i \) will grow at a double-exponential rate, giving rise to the phenomenon of \textit{combinatorial explosion}.

The takeaway is simple but powerful.  
\textit{Even a small increase in accessible knowledge leads to an extremely large increase in innovation potential.}  

That’s why AI, which helps researchers find, filter, and recombine knowledge, can have such a huge effect. If AI boosts the effective \( \phi \), researchers can access more knowledge, and the number of useful combinations explodes exponentially.

This phenomenon, \textit{combinatorial explosion}, explains why AI may fundamentally change the paradigm of scientific discovery.

The contribution of much of the argumentation here is to link the combinatorial potential of innovation directly to the moral and ethical foundations articulated in EMT. As discussed, according to EMT, the purpose of scientific and economic activity is not primarily growth for its own sake, but the alignment of knowledge production with a dynamically evolving frontier of human experiential needs. Extending Callaghan's (2025a; 2025b) development of EMT, this paper formalises its combinatorial foundations using a mathematical framework grounded in knowledge recombination.

Having incorporated into our discussions the above model of combinatorial growth proposed by Agrawal et al. (2018), we now extend prior discussions by demonstrating how science, understood as an industry, is subject to AI-driven acceleration through both product and process innovation channels. In doing so, we derive a logically consistent and mathematically defensible structure that positions AI as a necessary condition for achieving EMT’s vision of a post-scarcity economy oriented toward human flourishing.

We provide formal definitions, derive key results, and offer proofs that demonstrate how AI can navigate and make tractable, the combinatorial explosion inherent in the modern innovation landscape. Through this, we integrate ethical theory, innovation economics, and computational capacity into a unified framework for understanding science in the age of AI.

\section{Science as an Industry: Price Signals and Innovation Response}

Science is often treated as epistemically unique, governed by norms of disinterested inquiry, professional autonomy, and methodological rigor rather than by price mechanisms or market incentives. While these distinctions are culturally and philosophically important, they obscure a critical economic insight, that science, like any other system of production, operates within a resource-constrained environment and responds to the relative costs of its inputs. As ideation becomes less costly through AI-enabled automation, scientific production becomes increasingly responsive to these cost dynamics. This section reframes science as an innovation-responsive industry, one whose structure and tempo evolve in line with shifts in the cost of discovery.

In traditional settings, the high cost of ideation imposes strict constraints on what science can investigate. Research agendas are shaped by funding scarcity, long grant cycles, and the difficulty of recruiting skilled labor to cognitively demanding tasks. 

The institutional architecture of science, including peer review, publication norms, and the broader paradigm of rejection, emerged primarily as a mechanism to manage scarcity. Historically, production capacity was limited, for example, by journal page budgets and editorial resources, which constrained how many submissions could be accepted. Rejection thus functioned as a rationing device aligned with the economics of scarcity. Under these conditions, science operated much like a high-fixed-cost industry, characterised by slow throughput, uncertain outputs, and tightly controlled access to formal channels of dissemination.

However, the emergence of general-purpose AI systems, capable of literature synthesis, hypothesis generation, experimental design, and even autonomous knowledge construction, changes this dynamic. As modeled in the previous section, ideation costs now decline as a function of AI capability. This shift transforms the marginal cost curve of scientific production and creates a new set of optimisation conditions for institutions, funders, and research actors. 

As AI reduces the marginal cost of ideation, scientific output becomes increasingly elastic with respect to ideation input. In traditional models, the principal constraint was human cognitive bandwidth and the high cost of generating new, valuable ideas. AI fundamentally alters this dynamic, not only accelerating idea generation but also reducing epistemic noise, as anticipated by Weitzman (1998). This collapse in ideation friction creates the illusion that traditional constraints remain in place, when in fact many are dissolving. 

As AI dramatically reduces the marginal cost of ideation, scientific output becomes increasingly elastic with respect to ideation input. Historically, knowledge production was constrained by human cognitive limits, high costs of exploration, and epistemic noise. AI changes this fundamentally. As Weitzman (1998) anticipated, AI reduces ideational noise, making the discovery process more navigable and scalable. The result is not merely a shift in bottlenecks, but the removal of one of science's most persistent constraints, the high-friction nature of idea generation.

However, while the technical and epistemic barriers are collapsing, legacy institutional and cultural structures continue to simulate scarcity. Systems of peer review, prestige-based publishing, and institutional gatekeeping were designed to ration limited production capacity, and remain largely calibrated to a world of constrained ideation. As such, these systems often create the illusion that previous bottlenecks still exist. In reality, AI has rendered many of these constraints obsolete, but their institutional inertia maintains the appearance of limitation.

This disjuncture between post-scarcity technical capability and scarcity-based institutional design represents a fundamental tension in the contemporary scientific system. As we transition toward an ideation-abundant economy, new imperatives emerge, not from limits on cognition or computation, but from the need to align abundant ideation with ethical, social, and experiential priorities. 

\textit{For example, according to EMT's logics, the blind maximisation of production, such as growth measured solely by GDP, follows exponential logics that can lead to catastrophic outcomes}, including ecological collapse, resource exhaustion, technological destabilisation, and even human extinction. 

We argue that when growth is decoupled from human experiential well-being, it risks becoming directionless, extractive, and self-undermining. EMT provides a necessary corrective by offering an anchoring framework that aligns growth with the dynamic frontier of human experiential needs. Rather than treating growth as an end in itself, EMT reorients innovation, productivity, and economic expansion toward human flourishing, psychological depth, and existential alignment. By embedding ethical and experiential considerations directly into the objectives of growth, EMT offers a pathway to mitigate existential risks, avoid misaligned technological trajectories, and ensure safety across environmental, societal, and cognitive domains.

This is precisely the challenge addressed by EMT, which reframes scientific progress not as a function of scarcity, but as a process of aligning knowledge production with the evolving frontier of human needs.

As discussed, the collapse in the cost of ideation carries radical implications because it overturns core assumptions that have underpinned economics and social science for centuries. When the generation, recombination, and testing of ideas becomes exponentially cheaper, the foundational dynamics of innovation change. As explained, most existing models were built for a world in which ideation was scarce, costly, and slow. 

Continuing to operate with these outdated frameworks means misallocating resources, misunderstanding causal forces, and misreading signals at every level of policy and enterprise. This may result in inefficiency, perhaps even systemic failure. Catastrophic outcomes of this may include rising inequality, institutional breakdown, mass technological unemployment, misaligned AI deployment, and the erosion of democratic and epistemic legitimacy itself. 

EMT therefore protects by realigning economic theory with experiential needs of human beings. It replaces inert GDP metrics with a dynamic map of human fulfillment, providing a normative and analytical framework that is both empirically tractable and ethically grounded. In doing so, EMT enables societies to harness the unprecedented generativity of this new era, without falling prey to its disorienting risks.

We rely on the scientific system to guide us and avoid these kinds of risks. In terms of science, itself, treating it as an industry subject to economic incentives allows us to apply familiar theoretical tools. 

Under conditions of declining marginal cost, we expect substitution effects. for example, institutions may find it easier to reallocate resources away from traditional, labor-intensive ideation toward AI-augmented systems. We also expect scale effects. As the fixed costs of initiating complex research projects fall, more actors (including smaller labs and nontraditional institutions) enter the scientific marketplace. In short, science becomes more modular, decentralised, and contestable.

Importantly, the industrial analogy does not simplistically reduce science to a commodified service. Rather, it enables us to model how shifts in the relative cost of ideation reshape the structure of innovation itself. For example, in fields such as proteomics and synthetic biology, formerly intractable problem spaces are now increasingly navigable due to AI-enabled inference. The cost structure of these domains has changed, and with it, the feasible set of questions science can afford to answer. Science is no longer limited by what can be imagined through human intuition alone, but is increasingly shaped by what can be efficiently discovered through computational search.

This shift also highlights the importance of innovation responsiveness, the degree to which the scientific system reallocates effort in response to cost reductions. An industry with high responsiveness will rapidly shift resources toward tractable problem domains, realigning its production frontier. A sluggish or institutionally path-dependent system may fail to do so, resulting in misallocated attention, epistemic waste, or innovation asymmetries. Understanding science as an industry thus requires us to model not only the costs of ideation, but also the frictions and feedbacks that condition institutional responsiveness to those costs.

The reframing of science as an innovation-responsive industry prepares the ground for a deeper theoretical synthesis. In the next section, we formalise how ideation cost collapse interacts with the dual innovation pathways of process improvement and product enhancement, building on the foundational models of Romer (1990) and Grossman and Helpman (1991) to model science’s transformation as both a technological and economic process.

To fully capture the implications of collapsing ideation costs, we must treat science not only as an isolated domain of epistemic virtue, but also as an industry, subject to incentives, resource constraints, and responsiveness to technological change. Like any industry, science exhibits both product and process innovation, reacts to price signals, and reconfigures itself when its cost structures shift.

Historically, as discussed, science operated under conditions of extreme ideation scarcity. The difficulty and costliness of discovery justified hierarchical labor structures, long publication lags, and prestige-based capital allocation. However, as ideation becomes increasingly automatable, the scientific enterprise begins to exhibit the characteristics of a high-elasticity industry, responsive to marginal productivity changes in both labor and capital, and particularly sensitive to technological substitution by AI.

We therefore model science as a dual-input production sector. The output is knowledge \( K \), which enters downstream sectors as both a productivity booster and a consumption good. Inputs include human cognitive labor \( L_s \) and artificial ideation capacity \( A(t) \), subject to a combined production function:
\[
K_t = F(L_s, A(t)) = \beta_1 L_s^{\theta} + \beta_2 A(t)^{1 - \theta}
\]
where \( \theta \in [0,1] \) governs substitutability, and \( \beta_1, \beta_2 \) capture respective productivity.

The responsiveness of scientific output to \( A(t) \) increases as ideation shifts from tacit to codifiable knowledge. This triggers a cascading transformation in scientific labor markets, education pipelines, and journal gatekeeping structures. As the marginal cost of ideation falls, scientific effort reallocates from low-yield search toward high-leverage functions, such as validation, integration, and alignment with social priorities (such as advocated by EMT).

In simple terms, here we model science as something that produces knowledge using humans and AI. As AI gets better, it does more of the idea work, and the job of human scientists shifts toward checking, refining, and applying those ideas. This will change not only science jobs, but also education, journals, and how society guides research.

Crucially, this model enables science to be situated within standard economic frameworks of industry evolution. As with agriculture, manufacturing, and services before it, science transitions through phases of labor intensity, mechanisation, and automation. The uniqueness of the current moment lies in the epistemic nature of what is being automated, such as idea formation, hypothesis generation, and recombination of knowledge, which are functions once considered irreducibly human.

By treating science as an industry, we clarify the mechanisms through which AI-induced ideation cost collapse reverberates through knowledge systems. This sets the stage for our next task, unifying the respective process and product innovation models of Romer (1990) and Grossman and Helpman (1991) into a single framework that might be useful in predicting or better understanding science's responsiveness to declining ideation costs.

\section{Growth Unification in Science: Modelling Ideation Cost Decline via Process and Product Innovation}

Understanding the transformation of science into an ideation-responsive industry demands a theoretical foundation that can model both the generation of novel knowledge and its translation into useful products and services. Traditional growth theory provides two complementary architectures, Romer’s (1990) model of process innovation and Grossman and Helpman’s (1991) quality ladder framework for product innovation. We follow, and build on, Acemoglu's (2009) discussions of these as fundamental frameworks relating to process and product innovation, respectively. 

Each captures a distinct axis of progress, efficiency and variety, respectively, and together they offer a tractable platform for formalising the dynamics of ideation in the transition to the post science regime. In this section, we present each model and then synthesise them into a unified innovation function responsive to AI-induced declines in ideation cost.

\subsection{Romer's Endogenous Growth Model and AI's Lowering of Ideation Costs}

To understand how AI may transform economic growth, we begin by examining the logic of Romer's (1990) endogenous growth model, which builds on the standard Cobb-Douglas formulation often used in neoclassical growth theory.

Romer’s seminal contribution to endogenous growth theory formalised the idea that economic progress is driven by the accumulation of non-rival ideas that enhance the productivity of capital and labor. In his framework, the economy produces final output $Y(t)$ using a continuum of intermediate goods, whose productivity depends on the stock of ideas or technologies $A(t)$ available at time $t$. A representative production function is:

\begin{equation}
Y(t) = A(t) K(t)^\alpha L(t)^{1 - \alpha},
\end{equation}

where all variables are functions of time \( t \), reflecting their dynamic nature in a growing economy. The components of the function are defined as follows:

\begin{itemize}
  \item \( Y(t) \): \ Total output or gross domestic product (GDP) at time \( t \). This represents the aggregate quantity of goods and services produced in the economy.
  
  \item \( A(t) \): \ Total factor productivity (TFP) at time \( t \). This captures the efficiency with which labor and capital are used. It reflects the level of technology, knowledge, institutional quality, and other factors that enhance the effectiveness of input use. In many models, \( A(t) \) grows exogenously over time.
  
  \item \( K(t) \): \ Physical capital stock at time \( t \), such as machinery, buildings, tools, and infrastructure used in the production process. This is accumulated through investment and depreciates over time.
  
  \item \( L(t) \): \ Labor input at time \( t \), typically measured in effective labor units. This can include both the number of workers and their productivity, depending on whether human capital is incorporated.
  
  \item \( \alpha \in (0,1) \): \ Capital's share of output, also known as the output elasticity of capital. It measures the proportion of output that is attributed to capital input, with the remaining share \( 1 - \alpha \) attributed to labor. The parameter reflects the degree of returns to each input under constant returns to scale.
\end{itemize}

\noindent
This function assumes: (1) Constant returns to scale, whereby doubling both capital and labor doubles output. (2) Diminishing marginal returns, as increasing one input while holding the other fixed leads to smaller and smaller increases in output. (3) Endogenous technological progress, with the term \( A(t) \) included within the model, and typically assumed to be driven by factors as scientific discovery or institutional improvement. Hence, Romer's (1990) approach is described as a foundational model of endogenous growth theory.

In contrast, long-run growth in many traditional exogenous growth models, such as the Solow (1956) framework, depends primarily on the externally determined progression of \( A(t) \). Technological improvement is treated as an autonomous process, rather than the result of intentional R\&D investments as purposeful innovation, policy intervention, or design creation within the model itself.

The accumulation of $A(t)$ in the Romer model is governed by:

\begin{equation}
\dot{A}(t) = \delta A(t)^\phi L_A(t),
\end{equation}

where $L_A(t)$ is the labor allocated to research, $\delta$ is a productivity parameter, and $\phi$ reflects the returns to knowledge accumulation. If $\phi = 1$, the model implies scale effects; if $0 < \phi < 1$, the model supports semi-endogenous growth (as in Jones, 1995).

Romer’s insight is that process innovation, as improvements in the methods of production, can be endogenised through investment in idea generation. In our framework, this mechanism maps directly onto the AI-enhanced collapse of ideation cost, in that as $C(t)$ declines, the marginal productivity of $L_A(t)$ increases, accelerating the accumulation of $A(t)$ and shifting the growth trajectory of the economy.

Romer's endogenous growth model therefore emphasises technological progress as driven by knowledge accumulation, which enhances the productivity of capital and labor. In his formulation, new designs (or blueprints) increase the variety of intermediate goods, improving production efficiency.

Extending this model, Romer (1990) explicitly treats technology as an \emph{endogenous} outcome of economic activity, particularly research and development (R\&D). Instead of assuming \( A \) improves over time, Romer models how new technological \textit{designs} or \textit{blueprints} are created within the economy through intentional investment in knowledge creation.

Hence, drawing on and summarising Romer's ideas, the production function becomes:

\[
Y = L^{1 - \alpha} \left( \int_0^A x_i^\alpha \, di \right)
\]

Where:
\begin{itemize}
  \item \( Y \) is total output.
  \item \( L \) is labor used in final goods production.
  \item \( x_i \) represents the amount of intermediate input of type \( i \), with \( i \in [0, A] \).
  \item \( A \) is the number of available designs or types of intermediate goods (i.e., the result of ideation).
  \item The integral sums over the continuum of input varieties \( x_i \), each produced using a unique design.
  \item \( \alpha \in (0,1) \) again measures the importance of intermediate goods in production.
\end{itemize}

This formulation introduces the concept of \textit{variety-driven growth}, whereby economic output increases not only from more capital and labor but also from a greater diversity of intermediate goods, each arising from innovation.

Including both the standard Cobb-Douglas and Romer's variety-based formulation helps us appreciate the conceptual shift introduced by endogenous growth theory. 
This makes Romer's approach especially suitable for analysing the role of AI. In traditional models, AI might simply raise the value of \( A \), treating it as an improved ``black box." But in Romer's model, AI is expected to directly transform the process of innovation by (1) reducing the fixed cost of ideation, \( C(t) \), (2) automating the generation of new designs, and (3) increasing the rate at which the stock of designs \( A \) expands.

Thus, the sensitivity of growth to the cost and speed of idea generation becomes central. As AI decreases \( C(t) \), the economy can afford to generate more designs, increasing \( A \) and, through the integral term, raising output \( Y \) in a self-reinforcing cycle.

These models reveal specifically some channels through which AI redefines the innovation process. AI, by lowering the cost of creating new designs, may supercharge growth, and it is precisely this transition, from a world where ideas are expensive and scarce, to one where AI makes them cheap and abundant, that forms the basis of the post science paradigm.

\subsection{Grossman and Helpman’s Quality Ladder}

Whereas Romer's (1990) model may be interpreted as emphasising process innovation, as improvements in the efficiency of production using existing inputs, Grossman and Helpman (1991) place greater emphasis on product innovation through their quality ladder framework, where growth arises from the introduction of higher-quality intermediate goods that replace older ones.

In their model, firms compete to improve the quality of differentiated goods, each occupying a position on a ladder of increasing sophistication. Innovation occurs when a firm successfully develops a higher-quality version of an existing product, displacing the incumbent and capturing temporary monopoly profits.

Drawing from their ideas, let $q_i(t)$ denote the quality of product $i$ at time $t$. The aggregate quality index $Q(t)$ across all varieties can be written as:

\begin{equation}
Q(t) = \int_0^1 q_i(t) \, di,
\end{equation}

and final output is a function of both quantity and quality:

\begin{equation}
Y(t) = F(Q(t), K(t), L(t)).
\end{equation}

Innovation occurs when labor $L_Q(t)$ is allocated to product-line R\&D, improving the quality $q_i(t)$ with probability proportional to $L_Q(t)$. The innovation process is stochastic, and the model captures the Schumpeterian logic of creative destruction.

In the post science regime, where AI accelerates design, testing, and deployment, the ladder of quality becomes easier to climb. As $C(t)$ falls, the rate at which firms can generate and implement higher-quality variants of existing goods increases, raising $Q(t)$ and shifting the frontier of consumer utility.

Accordingly, the quality ladder model captures product innovation through successive improvements in existing goods. Firms engage in R\&D to leapfrog competitors by creating higher-quality variants, with the output indexed by quality level \( q \). Growth results from the arrival of innovations that replace old varieties:
\[
Y = \sum_{j=1}^N q_j x_j
\]
where \( q_j \) increases stochastically with innovation and \( x_j \) is the quantity of good \( j \). The model emphasises creative destruction, where ideation drives product turnover. AI shifts the probability distribution of discovery upward by accelerating search in the design space and increasing the likelihood of breakthrough innovation, especially in combinatorially large fields. Creative destruction is another channel through which AI might act on science through its industry dynamics.  

\subsection{Creative Destruction: Aghion and Howitt's Schumpeterian Dynamics}

Aghion and Howitt (1992) formalise economic growth as the outcome of a sequence of quality-improving innovations that replace outdated technologies in a process known as \textit{creative destruction}. Growth arises from an R\&D race, whereby firms invest in research to invent higher-quality products, displacing existing ones. This dynamic is captured in a continuous-time framework with overlapping generations of innovations.

They relate the idea of quality ladders with firm value, which we interpret and summarise as follows.  

\medskip

\noindent Each product line experiences a stochastic arrival of innovations governed by a Poisson process with intensity \( \mu(t) \). A successful innovator displaces the incumbent and becomes the new monopolist, producing a good of quality \( q(t) = \lambda q_{\text{prev}}(t) \), where \( \lambda > 1 \) is the step-size of innovation.

Let \( V(t) \) denote the value of holding a monopoly on a product line at time \( t \), and let \( r \) be the interest rate. Then the incumbent’s Bellman equation is:

\begin{equation}
r V(t) = \pi(t) - \mu(t) V(t),
\end{equation}

\noindent where:
\begin{itemize}
  \item \( \pi(t) \): Flow profit from holding the monopoly,
  \item \( \mu(t) V(t) \): Expected loss of value due to creative destruction from a successful R\&D competitor.
\end{itemize}

The R\&D investment and innovation arrival rates are then related to each other. 

\medskip

A potential entrant decides whether to invest in R\&D by weighing the cost \( \psi \) of research against the expected gain from displacing the incumbent. The free-entry condition is:

\begin{equation}
\mu(t) V(t) = \psi,
\end{equation}

\medskip

\noindent implying that the innovation arrival rate \( \mu(t) \) is determined endogenously by R\&D investment incentives.

Accordingly, given that each successful innovation improves quality by a factor \( \lambda \), and arrives with intensity \( \mu(t) \), the aggregate growth rate of the economy is:

\begin{equation}
g(t) = \ln(\lambda) \cdot \mu(t).
\end{equation}

\medskip

Hence, this growth rate depends critically on (1) the magnitude of improvement per innovation (\( \lambda \)), the frequency of innovation (\( \mu(t) \)), and the structure of displacement incentives (\( V(t) \)).

We now model AI and the endogenous disruption of knowledge, reinterpreting these dynamics in the context of scientific knowledge production. Accordingly, AI reshapes the system as follows:

\begin{itemize}
  \item \textit{Increased Innovation Frequency (\( \mu_{\text{AI}}(t) \))}: AI reduces the cost of ideation \( \psi \), automates hypothesis generation, and accelerates discovery, thereby raising \( \mu(t) \).
  
  \item \textit{Enhanced Quality Jumps (\( \lambda_{\text{AI}} \))}: AI's ability to cross-validate, simulate, and extrapolate increases the epistemic reliability and depth of each ``innovation step", in effect, increasing \( \lambda \).
  
  \item \textit{Dynamic Revaluation of Knowledge Assets (\( V(t) \))}: Incumbent paradigms or theories may see declining epistemic value as AI systems generate superior alternatives, reducing the effective lifetime of traditional models.
  
  \item \textit{Endogenous Obsolescence Cost (\( \delta(t) \))}: Institutional rigidity or scientific conservatism, manifesting as resistance to abandoning falsified or disrupted knowledge, can be conceptualised as a rising obsolescence burden, \( \delta(t) \), which dampens realized growth even as AI accelerates \( \mu(t) \).
\end{itemize}

AI introduces a powerful force of epistemic creative destruction, potentially capable of challenging even the foundational assumptions of entire scientific fields through the collapse of costs of knowledge/ideation. As ideation accelerates, the responsiveness of scientific institutions becomes increasingly consequential. For example, traditional journal review systems, where submissions are tied to a single journal and can spend years in sequential review cycles, may inadvertently hinder the timely dissemination of transformative ideas. In this context, well-intentioned gatekeeping can lead to costly delays in societal progress. 

As AI continues to lower the cost and increase the speed of discovery, there may be a growing case for carefully considered reform, perhaps even regulatory guidance, to ensure that our knowledge dissemination infrastructure evolves in step with the technologies driving it. This is not a critique of journals per se, but an invitation to reimagine how best to serve a scientific ecosystem increasingly shaped by real-time, generative intelligence.

\subsubsection{Generalised Scientific Growth Equation}

To summarise some of these ideas in `other words', let us define an AI-augmented version of the Aghion–Howitt growth equation for the scientific domain:

\begin{equation}
g_{\text{science}}(t) = \ln(\lambda_{\text{AI}}) \cdot \mu_{\text{AI}}(t) - \delta(t),
\end{equation}

\noindent where:
\begin{itemize}
  \item \( \mu_{\text{AI}}(t) \): Arrival rate of AI-enabled paradigm-displacing discoveries,
  \item \( \lambda_{\text{AI}} \): Epistemic quality gain per scientific update,
  \item \( \delta(t) \): Cost of failing to update outdated knowledge systems (e.g., path dependence, institutional lock-in).
\end{itemize}

As discussed, the use of AI in scientific discovery does not merely increase the pace of innovation; it also necessitates the frequent abandonment of older, falsified, or low-utility models. Failure to discard outdated ideas leads to epistemic inertia, analogous to economic stagnation under high \( \delta \). Accordingly, in this setting we might expect that low-cost, high-frequency AI-enabled discovery process raises \( \mu_{\text{AI}}(t) \), that an epistemically open scientific culture keeps \( \delta(t) \) low, and that the result is rapid, self-renewing scientific progress.

AI thus acts not only as a general-purpose technology but as a catalyst of \textit{epistemic creative destruction}, forcing a real-time evolution of knowledge systems in which their survival depends on their adaptivity and falsifiability. 

Aghion and Howitt, interestingly, assume a Poisson process governs the arrival of innovations, characterised by a constant arrival rate over time. In what follows, we extend the framework to explore a wider class of arrival distributions, demonstrating that the transformative effects of AI hold robustly across distributional assumptions. This allows us to generalise theoretically beyond Poisson-based innovation dynamics and capture richer epistemic and economic implications of accelerated ideation.

\subsubsection{AI and Distribution-Agnostic Innovation Growth: Beyond the Poisson Assumption}

Aghion and Howitt (1992) model the arrival of innovations as a Poisson process with constant intensity \( \mu \), which governs the frequency of paradigm-displacing events. However, recent work by Jones (2023) provides a deeper mathematical foundation showing that the distribution of idea productivity need not follow any specific form such as Poisson, Pareto, or Weibull. Rather, what matters is the interaction between the number of idea draws \( K \) and the shape of the tail of the underlying productivity distribution \( F(\cdot) \).

Let \( Z_K \) denote the productivity of the best idea among \( K \) independent draws from a continuous distribution with tail \( \bar{F}(x) = 1 - F(x) \). According to Jones's logics:

\begin{equation}
\lim_{K \to \infty} \Pr\left( K \bar{F}(Z_K) \geq m \right) = e^{-m}, \quad \forall m > 0,
\end{equation}

\noindent which implies that \( K \bar{F}(Z_K) \) converges in distribution to an exponential random variable. This result is distribution-agnostic, requiring only that \( F(x) \) be continuous and strictly increasing. The key implication is:

\begin{equation}
\boxed{
K \bar{F}(Z_K) \xrightarrow{d} \text{Exponential}(1)
}
\end{equation}

\vspace{1em}

\noindent As \( K \) grows (via combinatorial expansion or AI acceleration), the maximum idea quality \( Z_K \) marches deterministically down the tail of \( F(x) \), and growth is guaranteed if:

\begin{equation}
Z_K \sim \bar{F}^{-1} \left( \frac{\epsilon}{K} \right),
\end{equation}

\noindent for some \( \epsilon \sim \text{Exp}(1) \).

\medskip

\noindent

Hypothetically, therefore, AI might lower the cost of ideation and evaluation, enabling a super-exponential increase in the number of viable idea draws \( K(t) \) over time. Even if idea productivity is drawn from:

\begin{itemize}
  \item Thin-tailed distributions (e.g., Exponential, Weibull, Normal),
  \item Log-Pareto or lognormal (intermediate-tailed),
  \item Or bounded distributions (e.g., Uniform),
\end{itemize}

\noindent the convergence of \( K \bar{F}(Z_K) \) ensures that the maximum-quality idea \( Z_K \) grows, provided that \( K(t) \to \infty \) sufficiently rapidly.

\vspace{1em}

This result implies that (1) \textit{AI guarantees progress} in science and technology not necessarily because of the shape of the idea space, but primarily because of the sheer acceleration in exploration, (2) \textit{institutional inertia} (reflected in a rising effective \( \delta(t) \), as in the Aghion–Howitt framework) can still bottleneck progress unless scientific gatekeeping adapts, and (3) \textit{journal systems, funding models, and peer review mechanisms} must evolve to avoid artificially capping \( K(t) \), lest they suppress the very process that drives exponential idea selection. 

The concern about artificially capping knowledge creation is not merely one of procedural fairness; it may involve systemic risk. If the knowledge excluded might contribute to mitigating existential threats, the opportunity cost could be substantial. As with a difference equation whose parameter is negative and greater than one, instability compounds over time. A system premised on exclusion, capping knowledge production through a paradigm of systemic rejection, may inadvertently create the conditions for its own disruption, particularly in an era when AI accelerates ideation and the demand for real-time, causally robust, safety-critical knowledge becomes increasingly urgent.

Such gatekeeping may risk suppressing methodological pluralism, delaying innovation, and reinforcing an artificial scarcity in the production and diffusion of actionable knowledge (Callaghan, 2024). 

This may represent a continuing legacy of an analogue-era scientific system, in which epistemic scarcity, driven by physical constraints on paper-based or hard-copy journal formats, necessitated extreme selectivity. 

Under such conditions, editorial gatekeeping evolved historically as a rationing mechanism, often privileging stylistic coherence or disciplinary orthodoxy over causal clarity or societal utility. 

However, in the digital age, where marginal publication costs approach zero and AI systems exponentially increase ideation capacity, such scarcity-based filtering is increasingly anachronistic. 

Unless recalibrated toward causal validity and epistemic relevance, these legacy structures risk becoming noise-preserving rather than signal-amplifying components of the scientific infrastructure.

If, however, the publishing system’s distinction between valuable knowledge and noise were grounded in causal inference and empirically verifiable criteria, then its selective function could be more clearly justified, on the grounds that it contributes to societal benefit through the reliable accumulation of knowledge. In such a system, rejection would not only be epistemically defensible but also fairer in practice, imposing less of a human toll on knowledge workers who currently operate within an environment shaped more by scarcity and opacity than by transparent, causal standards.

From this perspective, science should be stripped of its artistic or interpretivist components, where these add epistemic noise (Weitzman, 1998), as such noise undermines the signal extraction essential for causal identification and theory-building. 

In high-stakes domains, such as AI safety, where policy, innovation, and societal well-being depend on the precision of inference, this noise may impose significant social costs by distorting the knowledge production function. 

\textit{Journals that prioritise methodological aesthetics, stylistic novelty, or conceptual impressionism} risk obstructing discovery, particularly when they mask the underlying causal signal. By contrast, a \textit{stripped-down science}, identified as such, focused on testable models and causally relevant mechanisms, strengthens the defensibility of exclusive publication venues by enhancing their instrumental value in societal problem-solving. If publication filters align with signal-amplifying criteria, as described in the distribution-agnostic modeling above, their selectivity channels ideas more effectively toward high-leverage societal outcomes.

In short, the discussions above suggest AI ensures that creative destruction, and thus epistemic advancement, proceeds across \emph{all} scientific domains, regardless of how idea quality is distributed. The bottleneck is no longer ideation, but remains our willingness to let outdated knowledge be replaced.

\subsection{A Unified Framework of Innovation Channels and AI's Disruptive Role in the Science Industry}

Building on the preceding sections, we now model the impact of AI on science conceived as an industry. To do so, we draw on the three previously discussed major theoretical channels, namely those suggested by Romer (1990), Grossman and Helpman (1991), and Aghion and Howitt (1992), which together seem to account for much of the longstanding variance in how idea-driven influences have been theorised within industry-level innovation models. When applied to science itself as an industry, these channels offer a conceptual framework for anticipating the transformative influence of AI on the structure, productivity, and epistemology of knowledge production.

This approach seems appropriate because in the scientific domain, AI disrupts not only physical goods but also the foundational assumptions, paradigms, and research practices of knowledge systems. 

Accordingly, we propose that the science industry is subject to \textit{the same three fundamentally AI-relevant forces} as traditional innovation-based industries:
\begin{enumerate}
  \item \textbf{Process Innovation (Romer)}: AI lowers ideation cost, increasing the number of scientific methods, techniques, and tools.
  \item \textbf{Product Innovation (Grossman–Helpman)}: AI improves the quality and accuracy of scientific outputs, enabling better theories and models.
  \item \textbf{Creative Destruction (Aghion–Howitt)}: AI challenges obsolete knowledge structures, catalysing a transition to more accurate, dynamic, and real-time epistemologies.
\end{enumerate}

\noindent
A unified mathematical framework integrating all three channels enables predictive insight into how AI will transform not only economies, but also the epistemic foundations of human knowledge.

Accordingly, we propose a composite model in which the ideation function \( I(t) \) is associated with simultaneous changes in both process efficiency and product quality. Let:
\[
G(I(t)) = \lambda_1 \cdot \frac{dA(t)}{dt} + \lambda_2 \cdot \frac{dq(t)}{dt}
\]
where \( \lambda_1, \lambda_2 \) are weights that capture the marginal association between ideation intensity and the rates of process and product innovation, respectively. Full definitions of terms follow.

\vspace{1em}

\noindent\textbf{Definition of Terms}
\begin{itemize}
  \item \( G(I(t)) \): Composite innovation output generated from ideation at time \( t \), encompassing both process and product innovations.
  \item \( I(t) \): Ideation intensity at time \( t \), defined as the ratio of R\&D expenditure \( R(t) \) to the cost of ideation \( C(t) \).
  \item \( A(t) \): Process innovation or efficiency level at time \( t \); reflects improvements in how goods or services are produced.
  \item \( q(t) \): Product innovation or quality level at time \( t \); reflects improvements in the characteristics or performance of products.
  \item \( \lambda_1 \): Weight reflecting the marginal impact of ideation on process innovation.
  \item \( \lambda_2 \): Weight reflecting the marginal impact of ideation on product innovation.
  \item \( R(t) \): Total expenditure on research and development (R\&D) at time \( t \).
  \item \( C(t) \): Cost of generating a unit of ideation at time \( t \); decreases with AI-driven improvements.
  \item \( \frac{dA(t)}{dt} \): Rate of change (growth) of process innovation over time.
  \item \( \frac{dq(t)}{dt} \): Rate of change (growth) of product innovation over time.
  \item \( \frac{dA}{dR} \): Rate of process innovation improvement per unit of R\&D investment.
  \item \( \frac{dq}{dR} \): Rate of product innovation improvement per unit of R\&D investment.
  \item \( \frac{dR}{dC^{-1}} \): Change in R\&D intensity as the inverse of ideation cost increases; reflects how lowering cost enables more effective R\&D.
\end{itemize}

\vspace{1em}

 Accordingly, as AI lowers the cost of ideation \( C(t) \), we reparameterise the R\&D intensity function:

\vspace{1em}
 
\[
I(t) = \frac{R(t)}{C(t)} \quad \Rightarrow \quad G(I(t)) = \lambda_1 \cdot \frac{dA}{dR} \cdot \frac{dR}{dC^{-1}} + \lambda_2 \cdot \frac{dq}{dR} \cdot \frac{dR}{dC^{-1}}
\]

\vspace{1em}

Under the post science regime, where \( C(t) \to 0 \), \( \frac{1}{C(t)} \to \infty \), meaning that modest R\&D inputs can yield disproportionately large innovation outputs. This shift collapses the traditional trade-off between exploration and exploitation and leads to a regime of real-time, demand-responsive innovation.

To model these relationships under the post science paradigm, we propose a simple unified innovation function that integrates both Romerian process innovation and Grossman–Helpman product innovation into a single dynamic framework. Let $A(t)$ denote process-related ideation, and $Q(t)$ denote aggregate product quality. We define total output as:

\begin{equation}
Y(t) = A(t)^\varphi Q(t)^\gamma K(t)^\beta L(t)^{1-\beta},
\end{equation}

where $\varphi, \gamma > 0$ govern the output elasticities of process and product innovation. Both $A(t)$ and $Q(t)$ are functions of ideation, driven by AI-enhanced labor:

\begin{align}
\dot{A}(t) &= \delta_A L_A(t) \cdot \left( \frac{1 + \alpha_A A(t)}{C(t)} \right), \\
\dot{Q}(t) &= \delta_Q L_Q(t) \cdot \left( \frac{1 + \alpha_Q Q(t)}{C(t)} \right),
\end{align}

where $C(t)$ is the ideation cost as previously defined, and $\delta_A, \delta_Q$ are innovation productivity parameters for process and product improvements. The parameters $\alpha_A$ and $\alpha_Q$ represent feedback effects, where existing levels of $A(t)$ and $Q(t)$ increase the efficiency of further innovation, capturing increasing returns to ideation.

While the current model explicitly integrates Romerian process innovation and Grossman–Helpman product innovation, the Schumpeterian dynamics of Aghion and Howitt (1992) are acknowledged implicitly through the improving quality trajectory $Q(t)$, though without modelling displacement or incumbent obsolescence directly.

This formulation captures a key insight, that as AI reduces the cost of ideation, both axes of innovation, process and product, accelerate simultaneously, reinforcing each other. Process improvements raise the efficiency of transforming inputs into outputs, while product innovations enhance the value and diversity of those outputs. The result is a systemic transformation in the structure and rate of economic growth.

This unification underpins the transformation of ideation into an abundant resource. However, as the next section explains, this does not eliminate epistemic complexity, but reorients it. The next frontier lies in understanding how ideation transitions from hypothesis-driven deduction or abduction to demand-shaped innovation in real time.

In simple terms, this modelling explains how AI can supercharge economic growth by making it much cheaper and faster to generate new ideas. Traditionally, R\&D was expensive, and scientists had to choose between trying new risky processes or improving existing products. But with AI, coming up with new ideas becomes almost free, which means we no longer have to make that trade-off, we can do both at once.

Because we are modelling the science system itself, process innovation $A(t)$ can be interpreted as improvements in scientific methodology, such as AI-assisted literature synthesis, automated experimental design, causal inference engines, or real-time peer review augmentation, each enhancing the efficiency and precision of knowledge production. Product innovation $Q(t)$, in this context, refers to the quality of the outputs of science, such as sharper causal claims, novel empirical regularities, policy-relevant insights, and breakthroughs in areas such as health, energy, and AI alignment. As both process and product innovation improve, the science system becomes more capable of producing knowledge that aligns with the dynamically evolving needs and values described in EMT, an emergent outcome of the innovation process.

In summary, we have represented these dynamics using a function that suggests the more we invest in idea generation (ideation), the more process and product innovation we get. As AI lowers the cost of ideation, the same amount of R\&D effort produces much more innovation. We suggested that total economic output depends on how good we are at making things efficiently, how good our products are, and how much capital (machines, tools) and labor (workers) we have. The innovation functions for process and product improvements are boosted as the cost of ideation approaches zero. This leads to faster and more responsive innovation, where ideas can be developed in real time to meet what people and societies actually need. In short, AI doesn't just speed things up, it changes the entire logic of how innovation works.

A key contribution of our model is the suggestion that while the traditional scientific method relied on hypothesis testing, where researchers formed theories and then tested them, AI may now enable a shift to real-time, demand-driven innovation. In this emerging post science paradigm, the process is reversed. Instead of starting with abstract theories, AI systems can detect what people or societies need in real time and generate solutions immediately. 

This means that innovation becomes more responsive and direct, no longer limited by the slow and sequential logic of hypothesis formation and validation. As ideation becomes nearly costless and automated, innovation can flow continuously from real-world needs to real-world outcomes, without the traditional delays of scientific cycles.

\subsection{Epistemic Transformation: From Induction, Deduction, and Abduction to EMT-informed Exduction and Demanduction}

Extending the perspective of EMT (Callaghan, 2025a; 2025b), the dual innovation pathway considered in the preceding subsection feeds directly into the expansion of the experiential matrix, a dynamic, multidimensional space of human needs. 

It is also important to recognise that many of humanity’s most profound and higher-order needs, those not satisfiable through immediate goods or services, might, by definition, be accessible only through the long-run functioning and progress of the scientific system. These include not only traditional scientific goals such as the elimination of disease and the overcoming of chronic health challenges, but also emergent and future-facing objectives such as enhanced quality of life, longevity, radical life extension, and even the Kurzweilian aspiration toward human immortality. In this light, science serves not merely as a means of satisfying fixed needs, but as the very vehicle through which new categories of need become intelligible and actionable. As such, the scientific system is foundational to the discovery and pursuit of evolving higher-order human needs. both those we are aware of today, and those that will only reveal themselves through the continued unfolding of human knowledge.

Process innovation therefore expands the ability to deliver experiences efficiently, and product innovation diversifies and enriches the set of experiences available. As $C(t)$ declines, the system becomes increasingly capable of mapping output onto human needs in real time, provided that alignment mechanisms, such as ethical oversight, participatory governance, and value-sensitive design, are in place.

This synthesis reframes growth not as the blind expansion of output, but as the alignment of ideation with the full spectrum of human flourishing. In the previous sections, we discussed how the post-scientific growth regime enables a shift from uncertainty to risk in the epistemic structure of science, transforming the logic of discovery itself.

Accordingly, the collapse in the cost of ideation does not merely accelerate discovery but transforms the epistemological foundations of science itself. Traditionally, scientific progress followed a sequence of logical inference modes, such as deduction, induction, and abduction, each requiring high cognitive effort and constrained by information scarcity. 

Traditional modes of reasoning in science, such as induction, deduction, and abduction, have historically provided the primary frameworks for theory generation, hypothesis testing, and inference. However, in the emerging post science paradigm enabled by artificial intelligence, these modes are increasingly augmented, or in some cases supplanted, by novel forms of logic that better reflect real-time responsiveness to experiential and epistemic shifts. Two such logics, central to the structure of EMT, are \textit{exduction} and \textit{demanduction}.

\textit{Exduction} refers to the process by which latent or existing knowledge is selectively activated in response to evolving human experiential needs. It is neither purely deductive nor inductive; rather, it is an alignment-based reasoning mechanism that bridges the space between human demand and the current state of knowledge. 

Exduction is thus the ideational engine of the EMT framework. It relies on a form of cognitive-gravitational alignment, wherein the ``pull” of human needs activates dormant or underutilised pathways within the knowledge base. This process is increasingly performed or assisted by AI systems capable of recognising deep structural analogies and mapping needs to existing solutions in real time.

In contrast, \textit{demanduction} refers to the reverse but complementary logic. While exduction begins with human needs and draws out knowledge that aligns with them, demanduction starts from the side of systemic, institutional, or market demand conditions and works to generate ideational responses that optimise fit within those constraints.

Demanduction is therefore the economic or market-facing counterpart to exduction. It is particularly important in contexts where innovation must meet not only experiential need but also strategic feasibility, regulatory compatibility, or institutional adoption likelihood. It formalises how AI systems can simulate, anticipate, or even shape demand-side configurations and feed this information back into the ideation process.

In EMT, demanduction does not merely reflect consumer demand in a classical economic sense. Rather, it integrates broader patterns of social receptivity, institutional absorptive capacity, and political feasibility into the innovation calculus. As such, it contributes to a system of ideation that is not only aligned with needs but also \textit{viable} within the real-world constraints of distribution, legitimacy, and infrastructure.

\subsubsection{Two Logics, One System: Complementarity in EMT}

Together, exduction and demanduction represent the dual logic of innovation in the post-science paradigm. Both are mediated by AI systems capable of:

\begin{itemize}
    \item Interpreting and weighting complex streams of experiential data;
    \item Mapping those data against expansive, heterogeneous knowledge corpora; and
    \item Adjusting outputs dynamically in response to both upstream (needs) and downstream (demands) constraints.
\end{itemize}

In this context, EMT serves as the unifying framework that theorises the conditions under which alignment is achieved. Exduction is need-facing and epistemic; demanduction is constraint-facing and systemic. Their interaction defines the AI-enhanced innovation system that replaces linear, siloed models of knowledge production with real-time, feedback-rich ideation flows.

Scientific reasoning is traditionally structured around the three fundamental modes of inference deduction, induction, and abduction. 

\textit{Deduction} proceeds from general principles or rules to derive specific conclusions that must logically follow if the premises are true. For example, from the general rule that ``all humans are mortal" and the fact that ``Socrates is a human," we deduce that ``Socrates is mortal." 

\textit{Induction} generalises from specific observations to broader patterns or laws. For instance, observing that the sun has risen every day in recorded history might lead to the inductive generalisation that ``the sun rises daily." However, inductive conclusions are inherently probabilistic and remain vulnerable to exceptions.

\textit{Abduction}, as introduced by Peirce (1935), refers to the process of inferring the most plausible explanation from incomplete or surprising evidence. Unlike deduction and induction, abduction does not guarantee truth or probability but instead generates hypotheses that best explain observed phenomena. Abduction plays a central role in scientific discovery, especially under conditions of uncertainty, where the goal is to construct coherent explanatory models from partial or noisy data.

Whereas deduction remains foundational for internal logical consistency but cannot generate novelty, induction requires extensive and often costly observation before generalisation is possible, and it remains vulnerable to overfitting and confirmation bias. Abduction, often described as inference to the best explanation, is prized for its creative reach. but it is fragile under uncertainty, deeply reliant on intuition, and structurally unsuited to large, multi-causal problem spaces. Moreover, all three operate under the assumption that knowledge production is supply-driven, that hypotheses, models, and experiments emerge from within the scientific system and are only later tested for relevance or application. As ideation costs fall, this assumption breaks down. Science becomes responsive not only to internal logic but also to external needs, values, and preferences. The problem space of discovery expands, and so too must the epistemic logic we use to navigate it.

These classical modes rely on human insight under conditions of high uncertainty and low information throughput. The tacitness and costliness of ideation historically made these modes central to knowledge formation.

Each has played a role in the evolution of science, but each was developed in contexts where ideation was scarce, slow, and largely human-driven. The collapse in ideation costs wrought by AI makes this legacy epistemology incomplete. 

Hence, AI undermines the conditions that made these logics dominant. With access to vast data corpora, real-time inference engines, and generative capabilities, large-scale models like GPT-4 and AlphaFold2 (https://alphafold.ebi.ac.uk/) already seem to outperform human abduction in several domains. As AI systems grow more context-sensitive and purpose-aligned, the limitations of traditional epistemic modes become increasingly binding.

In contrast, exduction and demanduction represent the dual logic of EMT's explanation of innovation and ideation in an AI-mediated world, the mapping of needs to hypotheses as a cognitive-experiential process, and the corresponding activation of knowledge as a formal, often algorithmic, response mechanism. The former grounds the theory in philosophical logic and epistemic alignment; the latter enables its application in models of artificial ideation, real-time research generation, and post-scarcity system design. These logics are now formalised as follows.

\subsubsection{Formalising Exduction}

In formal terms, let $E_t$ represent a vector of human experiential needs at time $t$, and let $\mathcal{K}$ denote the existing knowledge base. Exduction is the mapping function:

\begin{equation}
\mathcal{X}(E_t, \mathcal{K}) \rightarrow H_t,
\end{equation}

where $H_t$ is the set of hypotheses or candidate solutions activated by the alignment between needs and knowledge. Exduction operates as a kind of epistemic inverse function, whereby rather than asking “What can we know?”, it asks “What do we need to know to achieve this experience?” It is both backwards-inductive and forward-aligned, structured to prioritise human flourishing under post-scarcity ideation conditions.

AI systems, equipped with access to comprehensive knowledge graphs, reinforcement learning from human feedback (RLHF), and value-sensitive design frameworks, can perform exduction by simulating need-solution mappings across high-dimensional knowledge spaces. 

AI will increasingly possess the capacity to analyse data from a 360-degree perspective, integrating an unprecedented range of data types, including not only structured and unstructured data, but also experiential, emotional, and tacit forms of human input. This multidimensional integration marks a defining characteristic of the emerging post science paradigm. By collapsing traditional barriers to data synthesis and interpretation, AI enables real-time research productivity, effectively compressing or even collapsing the temporal dimension of scientific inquiry itself.

Importantly, human input remains essential for specifying $E_t$, the shape and texture of what flourishing means, making exduction a hybrid epistemic logic.

Accordingly, exduction is a novel epistemic logic whereby knowledge is activated in real time based on experiential context. Rather than being discovered in the abstract, knowledge is summoned on demand, shaped by its alignment with human or institutional goals. 

Mathematically, this can also be represented as:
\[
K_t = f(E_t, \theta_t)
\]
where \( K_t \) is activated knowledge, \( E_t \) is the experiential input stream, and \( \theta_t \) is the alignment parameter, determined by EMT or other normative frameworks.

This principle is grounded in EMT, which reconceptualises knowledge not merely as a static stock but as a dynamic contributor to production, insofar as it advances the expanding frontier of human experiential needs. Within this framework, the science system is understood as an industry in its own right, one whose output is evaluated by its capacity to fulfil and evolve the matrix of human needs and experiences.

For example, in a clinical decision-support system, exductive reasoning retrieves relevant medical models in response to dynamic patient data, conditioned by the desired outcome (e.g., quality-of-life improvement rather than mere survival). Exduction thus moves beyond classical inference, in that it is less about finding the ``truth" and more about generating aligned, context-appropriate responses within the experiential matrix.

\subsubsection{Demanduction: Market-Aligned Ideation}

In a post science paradigm, scarcity has been overcome. The purpose of science changes. EMT suggests that science, like any other industry, should align its efforts at meeting a continually expanding frontier of human experiential needs, increasingly higher order needs (that have to date been difficult for markets to satisfy). 

As ideation costs collapse and real-time inference becomes feasible, science is increasingly shaped by demand-driven logic. As discussed, we define \textit{demanduction} as the mode in which market signals and societal needs dynamically steer research direction. In this model, ideation is not scarce but plentiful. The bottleneck shifts from ``finding ideas" to ``finding valuable problems worth solving."

AI accelerates this transformation by rapidly generating idea permutations, simulating outcomes and social responses, and filtering ideas through human-aligned utility functions.

Demanduction formalises this logic. Let \( U_h \) represent a human-centered utility function derived from EMT, and let \( \mathcal{I}(t) \) be the ideation stream from AI systems. The set of research directions \( \mathcal{D}(t) \) becomes:
\[
\mathcal{D}(t) = \arg\max_{i \in \mathcal{I}(t)} U_h(i)
\]

Here, ideation is continuous, and filtering is goal-oriented. The science system becomes not a search over an abstract space of unknowns, but a real-time optimisation over alignment space.

This answers the question of what are the research directions D(t) or the ideas produced by AI that are most valuable to humans, according to a human evaluation function. In other words, from all the ideas the AI is generating, which are the ones that humans find most useful.
\textit{Demanduction extends exduction into systems with market signals.} It is the process by which ideation is directed toward domains where latent or emergent demand is detected or anticipated. Unlike traditional R\&D, which speculates about future needs, demanductive systems constantly map the ideation frontier to real-time preference data, feedback loops, and revealed behavioral signals.

Formally, let $D_t$ represent a set of market-aligned signals, such as, but not limited to, price movements, search queries, purchase histories, health metrics, etc., and let $C(t)$ be the cost of ideation at time $t$. Demanduction is a feedback-optimisation function:

\begin{equation}
\mathcal{D}(D_t, C(t)) \rightarrow \arg\max_{I_t} \; U(E_t(I_t)),
\end{equation}

where $I_t$ is a portfolio of ideation investments and $U(E_t)$ is the utility of the resulting experiences. Demanduction thus connects the structure of innovation with the structure of needs as reflected in the marketplace, enabling a decentralised, demand-pulled model of scientific progress.

When paired with AI agents capable of interpreting both economic and experiential data, demanduction becomes a powerful mode of knowledge generation. Scientific research no longer operates in isolation from social valuation, but becomes dynamically aligned with it. In a world of zero ideation scarcity, the constraint on knowledge production is no longer epistemic capacity, but the alignment of that capacity with systems of meaning.

\vspace{1em}

The epistemic shift from deduction, induction, and abduction toward exduction and demanduction reflects a deeper transformation, from science as explanation to science as alignment. This transition is not merely methodological but institutional, restructuring not only how knowledge is generated but how it is validated, disseminated, and used.

It requires new systems of funding, governance, and ethical oversight to ensure that ideation remains aligned with collective wellbeing. In the final section, we formalise this alignment within the framework of EMT and offer design principles for building an innovation system suited to the post science era.

In sum, the epistemic function of science is inverted. Rather than discovering fixed truths through laborious inference, science in the post science paradigm generates truths that are responsive to experiential needs and governed by normative alignment. This has profound consequences for how research systems are designed, evaluated, and governed, which we explore in subsequent sections.

The unification of growth theory with analyses of AI's transformative impact on science, particularly its role in ushering a post-science paradigm, necessitates the formalisation of a logic that positions EMT as a comprehensive and unifying theoretical framework. EMT offers a means to align AI and human goals around the central imperative of satisfying an ever-expanding frontier of human experiential needs. Yet an even deeper and more foundational rationale must be formalised: \textit{Why EMT?}

At the core of human existence lies a hierarchy of needs, existential and otherwise, ranging from survival to the highest forms of self-actualisation, as articulated by Maslow and others. We now advance the argument that these needs are not merely passive states but exert a gravitational pull of their own. In other words, they act as an attracting force, a kind of experiential gravity, that over time, and especially within a post-scarcity, AI-enabled world, will increasingly shape economic and technological trajectories.

As discussed, we argue that at the core of human existence lies a hierarchy of needs, existential and otherwise, ranging from physical survival to the highest forms of self-actualisation and transcendence, as articulated by Maslow (1943; 1970) and others. Within the framework of EMT, we reconceptualise these needs not as static desires or subjective preferences, but as dynamically evolving forces that shape economic and technological systems.

We now advance the argument that these needs exert what may be called an \textit{experiential gravity}, an attracting force that influences the direction of ideation, resource allocation, institutional change, and innovation trajectories. Over time, and especially within a post-scarcity, AI-enabled world, this gravitational pull becomes increasingly dominant. As AI lowers the cost of ideation and expands the capacity for real-time alignment with experiential data, human needs begin to operate not only as passive endpoints but as active structuring forces in the economy.

This experiential gravitational dynamic underpins exduction and demanduction, the two foundational reasoning logics articulated in this section, which together define the alignment-based epistemology of EMT. While exduction is activated through experiential alignment, drawing upon human needs and existing knowledge to generate hypotheses aimed at fulfilling those needs, demanduction operates through AI-simulated models of acceptability and constraint recognition. It derives ideational direction from market and institutional demand structures, shaping proposals that are not only need-aligned but also viable within prevailing systemic conditions. Together, these twin logics reflect a shift from explanation-driven science to alignment-driven innovation, governed by the gravitational pull of evolving human needs.

\section{Formalising the Gravity Model of Experiential Needs}

To deepen the theoretical structure of EMT, we propose a formalised \textit{gravity model} of experiential alignment. This draws conceptual inspiration from gravity models in international trade and economic geography, where flows (e.g., of goods or people) are shaped by the ``mass'' of origin and destination and the ``distance'' between them.

Here, we reinterpret this gravitational metaphor to explain how \textit{unmet human needs exert a directional force} on innovation systems. Rather than being passive or exogenous, needs are treated as \textit{structural attractors} that shape ideation, production, and economic evolution, especially in a world where AI removes historic barriers to responsiveness.

Let \( N_i \) denote the \textit{intensity} of an unmet experiential need \( i \), and let \( D_{i,t} \) represent the \textit{epistemic or technological distance} between the current knowledge frontier and a viable solution to need \( i \). We define the gravitational influence of need \( i \) at time \( t \) as:

\begin{equation}
G_i(t) = \frac{N_i^\alpha}{D_{i,t}^\beta},
\end{equation}

where \( \alpha \) and \( \beta \) are elasticity parameters representing system sensitivity to need magnitude and proximity, respectively.

The total experiential gravitational field at time \( t \), \( G_t \), is then:

\begin{equation}
G_t = \sum_{i=1}^n \frac{N_i^\alpha}{D_{i,t}^\beta}.
\end{equation}

This gravitational field exerts force on two central mechanisms of EMT, on exduction whereby high-gravity needs activate aligned knowledge fragments, guiding ideation and solution formulation, and on demanduction, whereby institutional and market filters simulate how gravitational forces translate into feasible, prioritised innovation responses.

This formalisation marks a paradigmatic shift, whereby \textit{innovation is neither purely supply-driven (random discovery) nor demand-driven (consumer preferences), but gravity-driven}, a continuous reorientation of ideational and productive capacity toward emergent configurations of human needs. EMT thus redefines utility not in terms of consumption, but as alignment with an evolving topology of existential and experiential necessity.

\subsubsection*{Extending the Classical Gravity Equation to EMT}

The classical gravity model, used in predicting trade flows and migration, is based on Newton’s law:

\begin{equation}
F_{ij} = G \cdot \frac{M_i \cdot M_j}{D_{ij}^2},
\end{equation}

where:
\begin{itemize}
    \item \( F_{ij} \) is the flow between entities \( i \) and \( j \),
    \item \( M_i \) and \( M_j \) are the respective economic ``masses'',
    \item \( D_{ij} \) is the distance between them,
    \item \( G \) is a gravitational constant.
\end{itemize}

We now extend this structure to model \textit{flows of productive potential} in response to human needs:

\begin{equation}
F_{ij}(t) = G(t) \cdot \frac{N_i(t) \cdot P_j(t)}{D_{ij}(t)^2},
\end{equation}

where:
\begin{itemize}
    \item \( N_i(t) \) is the intensity of unmet experiential need \( i \) at time \( t \),
    \item \( P_j(t) \) is the productive potential of sector \( j \) to address that need,
    \item \( D_{ij}(t) \) is the epistemic or institutional distance between the need and its satisfaction,
    \item \( G(t) \) is a responsiveness coefficient (amplified by AI).
\end{itemize}

\noindent
In this extended gravity model of experiential needs, we include the term \( P_j(t) \), representing the productive potential of sector \( j \), to capture the system’s supply-side responsiveness. While \( N_i(t) \) quantifies the intensity of an unmet human need, needs alone do not generate outcomes, they require translation into feasible innovations or interventions. Sectors serve as loci of production, where ideational and technological resources can be mobilised to meet specific needs. The inclusion of \( P_j(t) \) allows the model to formalise how productive capacity is distributed across the economy in response to gravitational pulls exerted by various needs. The epistemic or institutional distance \( D_{ij}(t) \) reflects the friction between a need and a sector’s ability to address it. For example, the distance between a need for longevity and the biotech sector may be small, while the distance between social cohesion and urban infrastructure may be much larger. The coefficient \( G(t) \), amplified by AI, captures the responsiveness of the system as a whole, particularly in its ability to overcome frictions and reallocate attention or resources dynamically. This formulation enables the model to describe directional flows from human needs toward realisable solutions, mediated by the structure and adaptability of production systems.

Stronger needs generate stronger pull, but institutional or cognitive barriers can block flows. AI systems reduce these frictions, increasing system responsiveness and accelerating need-solution convergence.

In conventional economic systems, productive capacity is often trapped in \textit{path dependencies and inertial loops}, privileging output volume (e.g., GDP) over alignment with genuine human needs. This may result in \textit{blind growth}, misallocation, underemployment, and neglect of higher-order needs such as meaning, social cohesion, and self-actualisation.

The EMT gravity model explains how unmet needs create dynamic attractors that, with sufficient responsiveness (via AI), \textit{break the flywheel of misaligned growth}. AI not only amplifies ideational capacity but enables systems to respond to deeper, latent human demands. In this sense, EMT offers both a formal and normative theory, a gravitational model of how economic systems evolve toward alignment, and a framework for designing AI-augmented institutions capable of realising this alignment.

Accordingly, the gravitational force of unmet human needs is not speculative but structurally predictive. It acts as a deep attractor shaping the direction of ideation, production, and technological deployment. In a world increasingly governed by intelligent systems, this pull toward alignment is not optional, it is inevitable. EMT captures this inevitability formally, offering a model not just of preference satisfaction but of the fundamental \textit{directional logic of human-AI economic co-evolution}.

\section{The Flywheel Effect of Lock-Ins}
In traditional systems, output growth follows a formula like:
\begin{equation}
\dot{Y}(t) = f(K, L, A(t))
\end{equation}
\noindent
Where $K$ is capital, $L$ is labor, and $A(t)$ is technology. This model ignores $N_i(t)$, unmet needs, so production grows regardless of whether it is needed. We suggest this creates a ``flywheel effect" where economies produce blindly, reinforcing existing paths without re-evaluating purpose.

A powerful illustration of the risks inherent in unbounded, goal-misaligned growth comes from Bostrom's (2003) thought experiment of the ``paperclip maximiser." In this scenario, a superintelligent AI is given the seemingly innocuous directive to manufacture paperclips. However, due to the lack of alignment between its operational goal and broader human values, the AI begins converting all available matter, including human bodies and the Earth itself, into paperclips. This example is not merely science fiction. It represents a profound warning about systems driven by singular, unqualified maximisation functions. Analogously, GDP growth, when pursued blindly as an end in itself, risks becoming a form of institutionalised paperclip maximisation, optimising production and consumption without regard to the evolving, multidimensional structure of human needs. Within the EMT framework, this critique is central, whereby economic growth must be evaluated not by its volume, but by its alignment with the dynamic experiential matrix of human well-being. AI offers the potential to shift us from a paradigm of blind optimisation to one of intelligent alignment, but only if its objectives are anchored in a framework like EMT that recognises the full gravity of experiential needs.

Additionally, according to EMT (Callaghan, 2025a; 2025b), unemployment is, in principle, largely irrational if the dynamic experiential frontier of human needs is effectively infinite and continually expanding, especially as AI reveals and enables the fulfilment of previously unarticulated or unmet needs. 

EMT posits that even under conditions of full AI utilisation, full human employment remains both possible and necessary, as the matrix of human needs perpetually outpaces available productive capacity. 

The theory therefore attributes persistent unemployment not to technological displacement \textit{per se}, but to the structural flywheel effect modelled here, a self-reinforcing inertia rooted in legacy systems of production, institutional lock-ins, and misaligned growth objectives. Crucially, EMT holds that AI has the potential to mitigate this flywheel effect by enabling more precise alignment between productive capacity and the evolving structure of human needs, thereby restoring employment as an essential mechanism for experiential value creation. The primary role of AI is therefore alignment, or realignment of growth to better focus on unmet needs.

\subsection{EMT and AI Realignment}

To formalise these ideas further, we now introduce a new term capturing alignment with unmet needs. Let the production function be modified as follows:

\begin{equation}
\dot{Y}(t) = f\left(K(t), L(t), A(t), \sum_{i=1}^{n} F_{ij}(t)\right)
\end{equation}

\noindent
Here, \( \sum_{i=1}^{n} F_{ij}(t) \) represents the total flow of resources directed toward satisfying the \( j \)-th category of real human needs at time \( t \), aggregated across all need types \( i = 1, \dots, n \). AI systems enhance this alignment by detecting, prioritising, and reallocating productive capacity based on real-time data about what people need most.

Equation (35) focuses on the production dynamics within a single category of goods or services, indexed by \( j \), and models how flows of productive attention toward this sector arise from the full set of unmet human needs \( i = 1, \ldots, n \). The term \( \sum_{i=1}^n F_{ij}(t) \) thus captures the total gravitational pull exerted on sector \( j \) by all existing needs in the experiential matrix. This formulation allows for a granular analysis of alignment at the sectoral level, illustrating how responsive a given part of the economy is to evolving need structures. While the model currently isolates one sector \( j \), it can be generalised across all sectors by summing over \( j \) to recover an economy-wide production function. The sector-level framing enables targeted investigation of bottlenecks, misalignments, or accelerators in specific domains (e.g., healthcare, education, clean energy) where productive capacity may or may not align with the most urgent components of the human needs landscape.


Accordingly to this modelling, instead of blindly maximising output by ``making more things," the economy begins to better listen to the voice of human needs. In this model, AI functions as a hearing aid, translating diffuse experiential signals into actionable directions for production.


An alternative but complementary perspective is offered by defining a form of \textit{social potential energy} embedded in unmet needs. Let:

\begin{equation}
U(t) = \sum_{i=1}^{n} N_i(t)^2
\end{equation}

\noindent
where \( N_i(t) \) represents the intensity of unmet need \( i \) at time \( t \). The function \( U(t) \) captures the cumulative tension or pressure exerted by the experiential matrix on the production system. 
In Equation (36), the intensity of each unmet need \( N_i(t) \) is squared to reflect its nonlinear contribution to the system’s cumulative tension. This squaring serves multiple purposes. First, it disproportionately amplifies the influence of highly intense needs, capturing the intuition that extreme unmet needs exert a disproportionately stronger pull on attention and resources. Second, squaring guarantees that all contributions are non-negative, ensuring that the total potential energy \( U(t) \) is always non-negative and interpretable as a form of stored pressure. Finally, the squared form facilitates later use in gradient-based optimisation or dynamic models, where the derivative of \( N_i(t)^2 \) yields a smooth and interpretable force term proportional to the magnitude of need. In this way, the squared formulation aligns both mathematically and conceptually with the idea of social or experiential potential energy embedded within the structure of unmet human needs.


According to this model, minimising \( U(t) \) thus becomes a normative objective for AI-enhanced economic systems. Just as physical gravity minimises potential energy in mechanical systems, aligned production within the EMT framework seeks to reduce experiential potential energy by closing the gap between needs and outputs. In this paradigm, economic activity becomes an optimisation process grounded in alignment rather than arbitrary expansion.

In summary, this gravity-based model offers a new lens on growth, one that moves beyond GDP to frame economic performance as a function of how well we satisfy the full spectrum of human needs. EMT predicts that AI will eventually break the epistemic lock-ins of the GDP flywheel by enabling real-time alignment between productive capacity and the gravitational field of human experiential demand.

The reframing of economic output and innovation through the lens of experiential alignment lays the foundation for a deeper transformation in the structure of scientific activity itself. If unmet human needs exert a gravitational pull on production systems, and if AI enables real-time sensing and redirection of effort, then science itself ceases to be a slow, linear pipeline of discovery. Instead, it becomes a high-frequency, \textit{recursive} process of continuous alignment and optimisation. As ideation costs collapse under AI augmentation, the binding constraints in knowledge production shift, from information scarcity to decision bottlenecks, attention allocation, and feedback integration. In what follows, we develop a formal model of research productivity as a recursive optimisation problem, capturing how AI reconfigures the dynamics of inquiry, experimentation, and learning in real time.

\section{Recursive Optimisation and Real-Time Research Productivity}

The collapse of ideation costs fundamentally alters the temporal architecture of scientific discovery. Traditional research models, often linear, grant-driven, and slow to adapt, are poorly suited to environments where ideation can occur continuously and adaptively in response to unfolding information. 

As AI systems increasingly perform \textit{recursive} hypothesis testing, experimental design, and real-time inference, science itself becomes a \textit{dynamic stochastic process}. Accordingly, in this section, we frame scientific ideation as a recursive optimisation problem using Bellman principles, and we define a model of real-time innovation surplus that captures how path dependencies shape scientific productivity.

With ideation cost collapse, the economics of science shifts from a linear discovery pipeline to a recursive optimisation problem. In this new regime, ideation is no longer scarce but abundant, making attention, alignment, and feedback the primary constraints on research productivity. We therefore model scientific activity as a stochastic, recursive problem solvable in real time, capturing the dynamics of AI-augmented discovery and decision-making.

\subsection{A Bellman Equation for Research Productivity}

As ideation costs collapse and AI systems gain the ability to evaluate, generate, and test ideas in real time, the structure of scientific activity begins to resemble a recursive optimisation problem rather than a linear discovery process. To model this transformation, we introduce a Bellman equation framework for real-time research productivity.

In modelling the transformation of research under the post science paradigm, we adopt a Bellman equation framework rather than an optimal control approach. This choice is both methodological and conceptual. Bellman equations, which emerge from dynamic programming, are especially well-suited to systems characterised by recursive structure, decision uncertainty, and real-time adaptation, which are precisely the conditions under which we envision AI-augmented research operating.

The Bellman equation presented here is not intended for tractable solution, simulation, or empirical estimation. Rather, its purpose is heuristic and conceptual, serving as a symbolic formalisation to provoke deeper understanding of scientific discovery dynamics under the post science paradigm. By framing discovery as a sequential decision-making process involving state transitions, strategic actions, and stochastic innovation shocks, the equation provides a stylised lens through which policy makers, institutional stakeholders, and scientific leaders can conceptualise the recursive, uncertain, and intertemporal nature of ideation and knowledge generation. 

The inclusion of the discount factor $\beta$ signals the necessity of long-term thinking in research policy, while the stochastic term $\varepsilon_t$ reflects the inherent uncertainty and serendipity characteristic of the innovation process. 

In this context, the Bellman equation is deployed as a symbolic device, a form of ``stylistic mathematics", to reorient stakeholder attention toward the structural logic of recursive value creation in science. It is intended to shift thinking from static and linear planning models toward a dynamic and forward-looking understanding of research system behaviour, supporting strategic reflection and policy innovation rather than precise analytical solution.

In our model, research decisions are therefore made recursively at each time step, based on the current state of knowledge and institutional context, subject to stochastic innovation shocks. The Bellman formulation allows us to incorporate this uncertainty naturally via an expectation over future value, making it ideal for capturing the probabilistic and exploratory nature of scientific discovery in an AI-enabled environment.

By contrast, traditional optimal control theory is better suited to deterministic or continuous-time systems in which control paths are planned in advance and solved via Hamiltonian methods or Pontryagin’s Maximum Principle. 

While optimal control excels in modelling aggregate, trajectory-based systems such as classical growth models or optimal capital accumulation, it lacks the flexibility and recursive logic needed to represent adaptive, feedback-rich systems where decisions unfold incrementally based on updated information. Moreover, Bellman equations directly mirror the logic of reinforcement learning and AI policy functions, which are essential to the real-time ideation, evaluation, and adjustment processes described in EMT. 

Rather than identifying a single optimal path, the Bellman framework generates a policy, a mapping from each state to the best action, given the expected long-run value of future outcomes. This aligns naturally with the EMT vision of research as a continuous, need-aligned optimisation process, where marginal gains in experiential utility guide the evolution of scientific systems. Therefore, the Bellman equation not only offers mathematical tractability in the presence of uncertainty, but also provides a conceptual foundation for modelling intelligent alignment in dynamic knowledge ecosystems.

Applying the Bellman approach, let the state variable \( s_t \) represent the configuration of the research system at time \( t \), including:
\begin{itemize}
    \item The current stock of scientific knowledge,
    \item Institutional conditions (e.g., funding systems, peer review, IP regimes),
    \item Available computational and cognitive resources.
\end{itemize}

Let \( a_t \) denote a research action taken at time \( t \), such as launching a project, generating a hypothesis, publishing a paper, or updating a model. The reward function \( r(s_t, a_t) \) represents the value of this action, as measured in terms of its alignment with human needs or systemic utility (as defined by EMT or other normative frameworks).

We assume the system evolves according to the following transition function:
\[
s_{t+1} = f(s_t, a_t, \epsilon_t)
\]
\noindent
where \( \epsilon_t \) represents a stochastic innovation shock, capturing the inherent uncertainty and serendipity in discovery processes.

The goal is to maximise the expected cumulative value of research over time. This yields the following Bellman equation:

\begin{equation}
V(s_t) = \max_{a_t} \left[ r(s_t, a_t) + \beta \mathbb{E}_{\epsilon_t} \left[ V(s_{t+1}) \right] \right]
\end{equation}

\medskip

\noindent
Where:
\begin{itemize}
    \item \( V(s_t) \): The expected value of being in state \( s_t \) at time \( t \).
    \item \( a_t \): The action chosen at time \( t \).
    \item \( r(s_t, a_t) \): The immediate reward from taking action \( a_t \) in state \( s_t \).
    \item \( \beta \in (0,1) \): The discount factor, which reflects the value placed on future outcomes relative to present rewards.
    \item \( \mathbb{E}_{\epsilon_t}[V(s_{t+1})] \): The expected future value of the next state, accounting for uncertainty in innovation outcomes.
\end{itemize}

\medskip

\noindent

In simple terms, to understand the logic of this model, you might think of the research system as a game where, at each step, you know the current situation (your knowledge and context), and you get to choose what to do next, for instance to run an experiment, launch a project, or publish a result. Every choice gives you some benefit now (a reward), but it also changes the situation for the future, possibly opening new directions or opportunities.

The Bellman equation offers a mathematical way of saying: ``What’s the best next move I can make, not just for today’s benefit, but for the future payoff as well?” It evaluates each possible action by adding the immediate reward from doing it now, and the expected long-term benefit of where that action leads, including surprises.

When AI systems use this logic, they no longer just produce outputs like papers on a conveyor belt. Instead, they continuously adapt, experiment, and learn, always asking, ``What’s the next best action to take, given everything I’ve learned so far?”

\medskip

\noindent

This recursive framing captures a core transformation in how knowledge is produced. In traditional systems, research progress is evaluated based on discrete outcomes, such as individual papers or grants. But in an AI-enhanced, EMT-aligned world, the focus shifts to real-time responsiveness, continuous learning, and the cumulative alignment of actions with evolving human needs. The Bellman equation formalises this shift. It models research not as a static product, but as a dynamic process of intelligent adjustment within a system shaped by both uncertainty and ethical responsibility.

The inclusion of the Bellman model here is not intended merely as a technical suggestion for policymakers or practitioners. Rather, it serves a deeper theoretical and structural role within the logic of EMT and the emerging post-science paradigm. First and foremost, the Bellman formulation offers a formal representation of a profound epistemic shift, from static, linear conceptions of science as a pipeline of discrete outputs, to a recursive, real-time optimisation process in which each research action is evaluated by its marginal contribution to long-run, aligned value. In this sense, the Bellman equation functions as a conceptual scaffolding that captures how ideation, experimentation, and feedback are recursively structured in AI-augmented research systems. 

Second, the framework provides a modelling language for understanding how AI itself may increasingly operate within the research domain, by allocating attention, generating hypotheses, updating beliefs, and selecting actions based on real-time evaluation of expected future rewards. 
This renders the Bellman structure not just a metaphor, but a mechanism for anticipating and simulating the dynamics of intelligent knowledge production. 

Third, the Bellman equation as expressed here supports EMT's normative commitments by conceptually embedding experiential alignment directly into the reward function \( r(s_t, a_t) \), allowing the recursive process to prioritise not merely technical outputs, but those actions which most meaningfully advance the frontier of human needs. In this way, the framework operationalises EMT’s call for aligned productivity in a world of abundant ideation. 

Finally, although not its primary purpose, the Bellman logic also offers potential policy relevance. It can inspire institutional architectures capable of adaptively steering science policy, funding allocation, or innovation ecosystems based on continuously updated feedback loops. In summary, the use of the Bellman equation in this model is foundational to capturing the recursive, uncertain, value-aligned, and intelligent nature of research systems under AI, and thus central to the theoretical and practical contributions of the paper.

\subsection{Ideation Cost Innovation Surplus}

According to the EMT logics outlined in the previous section, a reduction in ideation cost might fundamentally alter the marginal productivity of each new research action. Let \( C_I(t) \) denote the cost of ideation, the resources (time, effort, computation, or attention) required to generate a new idea or hypothesis, at time \( t \). As AI capabilities increase, this cost trends toward zero: \( C_I(t) \to 0 \).

Let \( r(a_t) \) be the reward function, representing the utility or system benefit generated by taking research action \( a_t \) at time \( t \). Then, the \textit{ideation cost innovation surplus}, denoted \( S_t^{\mathrm{IC}} \), which measures the payoff from each unit of research effort relative to its ideation cost, is given by:

\begin{equation}
S_t^{\mathrm{IC}} = \frac{\partial r}{\partial a_t} \bigg/ C_I(t)
\end{equation}

\noindent
Where:
\begin{itemize}
    \item $S_t^{\mathrm{IC}}$ is the \textit{ideation cost innovation surplus} at time \( t \).
    \item \( \frac{\partial r}{\partial a_t} \) is the marginal reward from taking research action \( a_t \); that is, the additional value or utility generated by incrementally increasing investment in that action.
    \item \( C_I(t) \) is the cost of ideation at time \( t \).
    \item \( r(a_t) \) may be interpreted as a proxy for short-run, alignment-adjusted research utility.
\end{itemize}

\noindent
\textit{Note:} In this section, the reward function \( r(a_t) \) is interpreted in a simplified form, capturing the marginal value of a research action in isolation. This contrasts with its earlier appearance as \( r(s_t, a_t) \) in the Bellman formulation, where it represents long-term, state-dependent utility within a recursive decision framework. Here, we focus on short-run surplus effects driven by the decline in ideation cost, abstracting temporarily from full system dynamics.

\medskip

\noindent
In this model, as \( C_I(t) \to 0 \), the denominator approaches zero, which causes \( S_t \to \infty \), unless bounded by system-level constraints such as knowledge saturation or misalignment with human needs. This illustrates the risk of unbounded innovation in the absence of normative filters. \textit{If ideation becomes nearly free, the system could generate overwhelming quantities of poorly targeted, low-impact outputs.} Therefore, recursive optimisation must be embedded within an alignment framework, such as EMT, to ensure that innovation remains bounded, meaningful, and socially valuable.

\noindent

\medskip

\noindent

This section might be summarised in general terms as follows. Imagine a world where coming up with new ideas or discoveries used to be expensive and slow, whereby scientists had to read, experiment, and think deeply over long periods. But now, with AI, the ``cost" of generating ideas is collapsing. It’s as if AI can brainstorm thousands of solutions instantly.

This equation suggests that if it's getting cheaper to generate new ideas (because of AI), and each new idea still gives you value, then the ``innovation surplus", the benefit you get per unit of effort, explodes. That seems positive, but there’s a catch.

If you’re not careful, you’ll end up with a flood of ideas that don’t solve real problems. Like publishing millions of useless research papers or building products no one needs. That’s why systems like EMT are necessary, to ensure that this explosion of innovation remains \textit{aligned} with real human needs. AI doesn’t just need to be fast, it needs to be meaningful.

This formulation therefore highlights the critical need for governance and alignment mechanisms, such as those proposed by EMT, to ensure that AI-enabled ideation does not result in noise, waste, or systemic overload. Recursive optimisation must therefore operate within an alignment-aware framework, in which innovation surplus is not only maximised, but also meaningfully directed.

\subsection{Real-Time Innovation Surplus and Dynamic Path Dependency}

While the previous section defined the innovation surplus in terms of declining ideation costs, capturing the structural benefits of AI over time, we now turn to a distinct but related concept, the \textit{real-time innovation surplus}. This surplus emerges not from cost savings per se, but from the dynamic gains generated by AI’s alignment with current needs and reflexive responsiveness to the evolving knowledge frontier.

Accordingly, in the previous section, we defined ideation cost innovation surplus as a measure of the
payoff from each unit of research effort relative to its ideation cost. We now extend this to model the surplus value generated by innovations in science assoicated with the transition to the post science paradigm. 

Accordingly, we define real-time innovation surplus as the gap between the instantaneous return
to aligned ideation and the counterfactual return under traditional research methods. 

Building on the prior discussion of path dependency and recursive optimisation, we now formalise the concept of real-time innovation surplus and examine how dynamic path dependencies can emerge internally within AI-augmented innovation systems. While earlier sections emphasised how institutional lock-ins and policy choices shape long-run epistemic outcomes, this section focuses on the internal logic of recursive optimisation itself and how, if left unguided, it may reinforce suboptimal or misaligned innovation trajectories.

We now formally represent the definition of real-time innovation surplus \( S_t^{\mathrm{RT}} \), as the gap between the immediate return to aligned, AI-augmented ideation and the counterfactual return under traditional, non-adaptive research approaches as follows:

\begin{equation}
S_t^{\mathrm{RT}} = R(s_t, a_t^{AI}) - R(s_t, a_t^{legacy}),
\end{equation}

\noindent
Where:
\begin{itemize}
    \item $S_t^{\mathrm{RT}}$ denotes the real-time innovation surplus at time \( t \),
    \item \( R(s_t, a_t^{AI}) \) is the reward or utility derived from an AI-augmented research action, aligned in real time with human needs and system context,
    \item \( R(s_t, a_t^{legacy}) \) is the reward from a conventional or legacy-mode research action, typically slower, linear, and less adaptive,
    \item \( s_t \) represents the state of the innovation system at time \( t \), including knowledge, context, and constraints.
\end{itemize}

As AI capability \( A(t) \) and alignment quality \( \alpha(t) \) increase over time, the differential value of adaptive research actions grows, implying that the opportunity cost of persisting with legacy scientific methods rises sharply. This reflects EMT’s proposition that traditional ideation becomes increasingly inefficient in the presence of recursive, AI-driven alignment mechanisms. This suggests a growing opportunity cost of not transitioning into AI-aligned ideation regimes.

However, real-time optimisation introduces new risks. Systems may overfit to transient trends, lock into local optima, or reinforce undesirable path dependencies if short-term returns are misaligned with long-term societal goals. Recursive optimisation frameworks must therefore be paired with safeguards for epistemic diversity and alignment with evolving experiential needs.

These structural risks arising from real time optimisation of scientific activity may arise because AI systems learn and adapt recursively. There therefore exists the potential for \textit{dynamic path dependency}, lock-ins that emerge not from initial institutional design, but from the behaviour of optimisation itself. 

Systems that overfit to short-term feedback signals or that optimise within narrow objective functions may converge prematurely on local optima, reinforcing shallow impact, ideological bias, or marketable but misaligned outputs.

These risks may be magnified in the presence of high-frequency decision loops and poorly specified reward functions. Just as policy path dependence can entrench inequality or inefficiency, recursive learning systems, if inadequately guided, can entrench epistemic inertia or generate brittle knowledge systems. The very tools that allow systems to adapt faster can also cause them to lock in faster.

Therefore, real-time recursive optimisation must be paired with safeguards for epistemic diversity, normative oversight, and structural adaptability. EMT provides the evaluative architecture for this oversight, ensuring that short-term optimisation is not decoupled from the long-term evolution of the experiential matrix of human needs. Alignment must not be seen as a static target, but as a moving frontier that requires continuous recalibration.

\medskip

\noindent
In sum, this section extends the recursive alignment logic of EMT not only by showing how real time innovation surplus might be generated by AI, but also that the internal dynamics of optimisation, if left unchecked, can generate new forms of lock-in. Real-time innovation surplus offers a useful metric for tracking the transition to AI-enhanced research systems, while dynamic path dependency reveals why this transition must be governed by principles deeper than efficiency, namely alignment, inclusion, and epistemic resilience.

The decomposition of these potential influences may enable further decompositional analysis of mechanisms through which AI reshapes innovation systems. While ideation cost reductions build capacity over time, real-time responsiveness accelerates the application and relevance of new knowledge to the present moment. Their joint operation explains how AI simultaneously unlocks long-term innovation efficiencies and short-term innovation alignment advantages, ultimately transforming both the supply and the timing of knowledge production. We now turn our attention to path dependency and policy sensitivity in the scientific system. 

\subsection{Path Dependency and Policy Sensitivity}

Because scientific knowledge systems and credibility architectures evolve incrementally over time, path dependency plays a critical role in shaping long-run innovation trajectories. Drawing on seminal insights from W. Brian Arthur and others in the economics of increasing returns and historical lock-in, we recognise that small, early differences in institutional configurations or epistemic norms can compound into dramatically different outcomes over time. In the context of a recursive, AI-augmented post-science paradigm, these effects are magnified.

Let \( s_0 \) represent the initial state of the research system, its configuration of knowledge, institutions, technological infrastructure, and incentive structures at time zero. Let \( P(t) \) denote the policy vector at time \( t \), comprising parameters that govern the innovation environment, such as funding allocation mechanisms, tenure and promotion criteria, publication filters, peer review rules, and institutional governance structures. Let \( V(s_0, P(t)) \) be the long-run value function over scientific productivity or social utility, conditional on both the initial state and the evolving policy trajectory. Then the evolution of optimal actions and outcomes is said to be \textit{path-dependent} if:

\begin{equation}
\frac{\partial V(s_0, P(t))}{\partial P(t)} \neq 0
\end{equation}

\noindent
Where:
\begin{itemize}
    \item \( V(s_0, P(t)) \) is the long-term expected value of the innovation system, conditional on the initial configuration and evolving policy.
    \item \( P(t) \) is the vector of policy instruments or institutional choices in effect at time \( t \).
    \item \( s_0 \) is the initial configuration of the system, which conditions the sensitivity of future trajectories to present decisions.
    \item The expression \( \frac{\partial V(s_0, P(t))}{\partial P(t)} \neq 0 \) signifies that changes in policy today have lasting consequences on system value, particularly given the system’s starting point.
\end{itemize}

\medskip

\noindent
The expression $\frac{\partial V(s_0, P(t))}{\partial P(t)} \ne 0$ captures the concept of path dependence in the evolution of innovation systems. Here, as indicated, $V(s_0, P(t))$ represents the long-term expected value of the system, conditional on its initial configuration $s_0$ and the trajectory of policy choices $P(t)$ over time. The non-zero derivative indicates that even small changes in policy instruments or institutional settings at time $t$ can produce lasting effects on the overall value trajectory of the system. This is particularly significant given the influence of the starting point $s_0$, which may amplify or constrain the effects of subsequent policy changes. Importantly, this sensitivity can be either positive or negative, whereby early choices may lock the system into high-value trajectories (e.g., robust innovation ecosystems), or conversely, into suboptimal paths with limited adaptability. As such, the expression serves as a formal heuristic to alert stakeholders that present-day decisions shape the future in ways that are conditioned by both historical context and structural feedback effects. The concept of path dependence thus provides a valuable lens for understanding why early, strategic policy interventions matter in complex, adaptive systems such as science and innovation. A non-specialist explanation of this model follows. 

Imagine steering a large ship. The direction you choose in the early part of the journey can send you thousands of miles off course if you're not careful. The same is true for science systems. If we embed AI into research in the wrong way, say, by rewarding papers over problem-solving, or by prioritising narrow metrics instead of human need, those early design choices can lock in patterns that are very hard to undo later.

This equation suggests that the value of the entire future research system depends directly on the policies we choose now, and on the starting point from which those policies are applied.

Even small differences in early funding rules, incentive structures, or data curation practices can echo across decades, amplifying biases or narrowing the scope of inquiry. 

Because scientific productivity in the AI era operates as a recursive, feedback-driven process, these early choices act as leverage points, and, once embedded, they shape what is seen, what is rewarded, and what is possible. That is what path dependence implies.

\medskip

\noindent
In the EMT framework, this insight becomes especially urgent. The cost of ideation is collapsing, which means the bottleneck is no longer the generation of ideas, but the alignment of those ideas with real human needs. The recursive nature of innovation means that small misalignments can compound over time into systemic distortions.

Therefore, the transformation to post science should begin with policies that prioritise experiential alignment, ethical governance, and equity of access, not as afterthoughts, but as foundational design principles. In this way, the Bellman structure and EMT logic converge to show that scientific progress is not policy-neutral, it is recursively shaped by the values we encode into its architecture from the very beginning. At this point, we close off our modelling of recursive systems by offering a cybernetic model of experiential governance.

\subsection{Recursive Alignment Feedback in AI Systems: A Cybernetic Model for Experiential Governance}

As the cost of ideation collapses and AI systems become increasingly capable of generating, evaluating, and synthesising knowledge, the challenge of aligning ideation with the evolving structure of human needs becomes paramount. In this context, alignment is not a static optimisation problem but a recursive, dynamic process. Drawing from Forrester’s (1961) system dynamics, Farmer’s adaptive learning models (Farmer and Foley, 2009; Farmer, 2024), and classical cybernetic control theory (Wiener, 1948), we propose a formal structure to model recursive alignment feedback within AI-driven knowledge systems.

\subsubsection*{System Variables and Alignment Error}

Let $O_t$ denote the ideation output vector at time $t$, representing the content, direction, and structure of ideas produced by AI systems. Let $E_t$ denote the experiential matrix at time $t$, capturing the dynamic constellation of human needs, values, and existential priorities. The instantaneous alignment error is defined as:

\begin{equation}
\epsilon_t = E_t - O_t
\end{equation}

This vector-valued error represents the gap between what ideation is producing and what is normatively needed for societal and civilisational flourishing.

\subsubsection*{First-Order Feedback: Cybernetic Control}

To model AI’s attempt to reduce this alignment error, we define a first-order control system in which the AI system adjusts its alignment signal $A_t$ in response to observed misalignment:

\begin{equation}
\frac{dA_t}{dt} = \gamma \cdot \epsilon_t
\end{equation}

Here, $\gamma > 0$ is a sensitivity parameter representing the rate at which AI responds to misalignment. When ideation outputs underperform in terms of alignment, $A_t$ increases to steer future outputs toward $E_t$.

\subsubsection*{Second-Order Dynamics: Adaptive Ideation Response}

Ideation output, however, is not fixed but evolves as a function of the control signal and stochastic perturbations:

\begin{equation}
\frac{dO_t}{dt} = \phi(A_t) + \eta_t
\end{equation}

Where $\phi(A_t)$ is a (potentially nonlinear) function mapping alignment effort into ideation shifts, and $\eta_t$ represents noise or innovation shocks. This captures Farmer’s insight that adaptive agents learn and change over time in response to feedback.

\subsubsection*{Third-Order Meta-Feedback: Learning from Learning}

To close the recursive loop, we introduce a meta-cybernetic layer in which the AI system adjusts its sensitivity $\gamma$ based on the rate of change in squared alignment error:

\begin{equation}
\frac{d\gamma}{dt} = \theta \cdot \frac{d\epsilon_t^2}{dt}
\end{equation}

Here, $\theta$ is the meta-learning rate. If alignment error is not reducing fast enough, $\gamma$ is increased, making the system more responsive; if overshooting occurs, $\gamma$ may decrease, preventing instability. This reflects a recursive self-tuning process in which AI learns how to learn alignment more effectively.

\subsubsection*{Systemic Implications for Governance and EMT}

The recursive cybernetic structure described above fundamentally reconceptualises alignment as an adaptive process of real-time governance rather than as a static calibration exercise. Within this framing, governance is no longer a top-down imposition of fixed goals onto scientific and technological trajectories, but rather a continuous modulation of ideation based on the dynamic interplay between human experiential needs and the outputs of AI systems. 

AI systems, under this formulation, must be equipped with internal feedback architectures that enable them to learn not just from data, but from their own evolving alignment performance. That is, the system must track and respond to its own history of adjustment success and failure, tuning its sensitivity parameters, such as $\gamma(t)$, in relation to whether alignment error is decreasing sufficiently over time.

Institutions, accordingly, must evolve toward foresight infrastructures capable of monitoring not just the volume and direction of ideation ($O_t$), but the velocity, precision, and meta-behaviour of alignment learning itself. This means that the metrics of institutional success must shift, from knowledge accumulation and publication volume, to recursive alignment sensitivity, ethical responsiveness, and impact on real human well-being as encoded in the experiential matrix ($E_t$). Within the context of EMT, this shift marks a deeper theoretical inversion. Optimisation now targets not knowledge quantity but minimisation of misalignment between ideation and the complex, multidimensional structure of human flourishing.

Moreover, as AI begins to recursively adjust its internal learning pathways in response to ethically weighted and socially deliberated targets, it takes on the characteristics of a moral-economic agent. Its behaviour is no longer governed solely by performance on a task or by external optimisation criteria, but by an embedded commitment to adjusting in relation to societal values. In this sense, alignment becomes an ongoing ethical negotiation embedded in the machine’s architecture, rather than a pre-defined constraint imposed from outside.

This has profound implications for policy. Governments and institutions no longer simply set one-time goals or funding priorities, but must continuously modulate the learning environment of AI by redefining $E_t$ (societal needs), monitoring $\epsilon_t^2$ (alignment error), and adjusting $\theta$ (meta-learning rate) to regulate how quickly and aggressively systems self-correct. Fail-safes, ethical override thresholds, and red-teaming mechanisms become essential elements of recursive governance.

In summary, this systemic framing under EMT does not merely advocate for better alignment but specifies a new governance paradigm in which feedback learning architectures, both human and artificial, are oriented around a shared goal, minimising the gap between what is ideated and what is experientially valuable. It is through this recursive loop that science, innovation, and policy converge into a cybernetically governed trajectory of human-aligned flourishing.

In sum, this framework extends the logic of EMT by embedding recursive alignment directly into the learning architecture of AI systems. Rather than viewing alignment as a fixed constraint, this model treats it as a continuously updating, cybernetically governed process, capable of evolving in tandem with the very needs it is meant to serve. 

We now turn our attention to the formal theoretical structure of EMT. EMT is built on the premise that the purpose of economic, scientific, and technological systems is not merely the maximisation of output, efficiency, or growth, but the recursive alignment of ideation and action with an evolving frontier of human experiential needs. This alignment includes both material and non-material dimensions, including health, meaning, self-actualisation, sustainability, autonomy, and collective flourishing. EMT reframes traditional optimisation problems, typically centred on production or consumption, into recursive learning systems governed by cybernetic feedback loops between ideation outputs and the experiential states of society. What follows is a consideration of formal aspects of the EMT framework, specifying its dynamic structure, recursive utility formulation, and implications for AI alignment, policy intervention, and systemic governance.

\section{Experiential Matrix Theory as a General Theoretical Framework for the Post Science Paradigm}

The paradigm shift triggered by AI and the collapse of ideation costs demands a theoretical framework capable of aligning this new epistemic surplus with human needs. EMT responds to this challenge by redefining growth and innovation in terms of alignment with the dynamically evolving frontier of human experiential needs.

While frameworks such as Romer’s process innovation model and Grossman–Helpman’s product innovation model continue to illuminate aspects of the transition from science to post-science, and have been integrated earlier for this purpose, EMT serves as the overarching framework that seeks to explain this transformation in full. 

EMT does not merely frame science as an industry subject to cost dynamics, but redefines the purpose of science itself by formally integrating ethical and moral alignment between human and AI activity. It provides a normative architecture that addresses not only how science functions, but what science is ultimately for. This reconceptualisation is both foundational and novel, differing from traditional economic models in important respects.

Traditional economic growth models, whether driven by process innovation (as in Romer’s endogenous growth model) or product innovation (as in Grossman–Helpman’s quality ladder), are structurally anchored in a paradigm of scarcity. These models optimise for output, efficiency, or quality improvements, often without interrogating whether the produced knowledge or products address the real and evolving needs of individuals and societies. EMT extends these frameworks by introducing a normative anchor, that growth is meaningful only when it satisfies the full spectrum of \textit{experiential} human needs.

\vspace{1em}
\noindent In the \textit{post science paradigm}, AI renders ideation increasingly abundant, scalable, and reflexive. This marks a structural rupture in the epistemic economy, transitioning science from a regime of \textit{Knightian uncertainty} to one of computable and probabilistic innovation. However, as stressed as a core concern of this paper, this epistemic inversion raises urgent questions of \textit{directionality}, asking what knowledge is worth producing, and for whom?

EMT addresses this by conceptualising the innovation process as a convergence mechanism, not simply a generator of output, but a dynamic reducer of the distance between \textit{what is needed} and \textit{what is possible}. As ideation costs collapse, the system’s ability to detect, model, and respond to unmet needs expands in real time. This transforms the ideational bottleneck into an alignment bottleneck, where the key constraint becomes not scarcity, but normative relevance.

To formalise these ideas, EMT offers a stylised form of an experiential matrix production function that integrates the infinite-dimensional phenomenology of human needs with a Cobb–Douglas structure to ensure tractability and analytical integration. It introduces the function \(\mathcal{E}(\cdot)\) which represents utility not over consumption alone, but also over aligned experiential fulfilment.

\vspace{1em}
\noindent EMT therefore unifies the economic and ethical dimensions of post-scarcity ideation. As previously discussed, it introduces two new epistemic logics unique to the post-science era, namely exduction, whereby knowledge is validated through resonance with real-time experiential needs, not only logical inference, and demanduction, as reflexive ideation steered by revealed or anticipated demand from within the experiential matrix.

Together, these logics replace classical modes of induction, abduction, and deduction in environments where knowledge generation is no longer the constraint, but alignment.

\vspace{1em}
\noindent In this formulation, EMT becomes not just a supplement to growth theory, but a general theory that integrates ethical purpose with algorithmic innovation. As we transition into the post-science era, EMT reframes science itself, not as a linear pipeline for discovering truth, but as a recursive, ethically grounded system for satisfying human needs through real-time, AI-enabled alignment. This section offers a brief overview of core EMT ideas as follows. For a more detailed introduction to EMT and its modelling see Callaghan (2025a; 2025b), from which the following logics are re-stated. 

Let the production function, as per Romer's (1990) endogenous growth model, be:
\begin{equation}
Y = A K^\alpha L^{1 - \alpha}, \quad \text{where } 0 < \alpha < 1
\end{equation}
where:
\begin{itemize}
  \item $Y$ is total output,
  \item $A$ is the stock of ideas/technology,
  \item $K$ is capital,
  \item $L$ is labor.
\end{itemize}

Let the left-hand side represent the \textit{Experiential Matrix}:
\begin{equation}
\mathcal{E}(x_1, x_2, ..., x_n, ..., x_\infty)
\end{equation}
where each $x_i$ corresponds to a phenomenological human need or dimension of experience (e.g., health, love, purpose, identity).

We define the foundational EMT equation as:
\begin{equation}
\mathcal{E}(x_1, x_2, ..., x_n, ..., x_\infty) = A K^\alpha L^{1 - \alpha}
\end{equation}

This equation suggests that economic output be oriented toward an unbounded and evolving constellation of human needs. Although the experiential matrix function $\mathcal{E}(\cdot)$ may defy closed-form solutions, it must nonetheless be recognised as a foundational structure, an ontological anchor, for any economic model that seeks to trace the trajectory of production systems toward a more ethical, human-centred future.

While Romer’s (1990) model envisions exponential growth propelled by the accumulation of non-rival ideas, and Weitzman (1998) even entertains the possibility of growth surpassing exponential rates, Jones (1995a, 1995b) offers a necessary corrective. By introducing diminishing returns to knowledge accumulation, Jones tempers the optimism of earlier models, aligning theoretical expectations more closely with observed empirical growth patterns. EMT absorbs this synthesis and reframes it, whereby it is not just the rate of growth that matters, but the direction, toward alignment with the full spectrum of human experiential value.

These foundational debates inform EMT’s predictions regarding the potential trajectory of exponential growth and underpin its core contention \textit{that growth increasingly unfolds within an intertwined physical and virtual experiential product and service matrix}. 

Accordingly, while the right-hand side of the present equation is inspired by Romer’s (1990) productivity function, EMT remains theoretically flexible and open to refinements that enhance both the precision and explanatory power of its modeling framework. Moreover, EMT's core argument is that when economic growth, as measured by gross domestic product (GDP) through aggregate production or consumption, remains the dominant policy objective, it risks incentivising outputs that are arbitrary or misaligned with real human needs.

Similarly, building on Arthur’s foundational work on increasing returns and path dependency (Arthur, 1989, 1990, 1993, 1994), EMT suggests that trajectories of change in production dynamics can become locked in through early investments, institutional reinforcement, and capability development, even when they cease to align with broader human welfare. From this perspective, these processes can widen the wedge between the blind maximisation of output and the fulfilment of complex, evolving human needs. Over time, reducing this wedge, by shifting production logics toward responsiveness to the experiential matrix, may offer a more meaningful pathway to human flourishing. This misalignment might be explained or modelled as follows. 
\vspace{1em}

Define a mapping operator:
\begin{equation}
\Phi(Y) = \hat{\mathcal{E}} \subseteq \mathcal{E}
\end{equation}
where $\Phi(Y)$ is the subset of the experiential matrix that is actually satisfied by current economic output.

The goal then over time would be to ensure that $\Phi(Y) \to \mathcal{E}$ as $t \to \infty$.


\vspace{1em}

\noindent Accordingly, we can represent the experiential matrix as follows. 

\begin{equation*}
\left( \begin{array}{cccc}
  x_{11} & x_{12} & \cdots & x_{1n} \\
  x_{21} & x_{22} & \cdots & x_{2n} \\
  \vdots & \vdots & \ddots & \vdots \\
  x_{m1} & x_{m2} & \cdots & x_{mn} \\
  \vdots & \vdots & \ddots & \vdots \\
  x_{\infty1} & x_{\infty2} & \cdots & x_{\infty n} 
\end{array} \right)
= A K^\alpha L^{1 - \alpha}
\end{equation*}

\begin{center}
\textit{Figure 1. Matrix Representation of the Experiential Matrix $\mathcal{E}$}
\end{center}

EMT offers implications for theory and policy. These include that theorists and policy makers must acknowledge that output alone might not guarantee human need satisfaction, that AI's contributions to productivity should also be evaluated based on its ability to help align $Y$ with $\mathcal{E}$, and that ethical AI and policy design must take the experiential matrix seriously, even if only partially mapped, or even if we cannot quantify it adequately yet. 

EMT offers a new way to conceive of economic purpose, shifting the focus from purely output maximisation to alignment with human needs. Even if the left-hand side of the equation remains under-defined, \textit{its presence disciplines the logic of economics, and even related social science theory and implications as they pertain to social benefit}. Future work should aim to quantify $\Phi(Y)$ more precisely and explore how AI can accelerate convergence toward $\mathcal{E}$.

EMT theorising recognises that people are not one-dimensional. Human well-being is not just a single number but a sum of many interactive needs being fulfilled, or left unfulfilled, over time. EMT models the full permutations of different kinds of human needs, such as the need for recognition, fairness, purpose, community, intellectual autonomy, and so on. EMT emphasises the persistent wedge between the satisfaction of needs within the current market system and the fulfillment of these diverse needs, including higher-order, humanist experiential needs.

\subsubsection*{Recursive Alignment and the Ultimate Purpose of Science}

To summarise ideas in this section, we formalise a central premise of EMT, that the purpose of science is the recursive alignment of knowledge production with the expanding frontier of human experiential needs, by introducing a dynamic, EMT-aligned utility function:

\begin{equation}
U_t = u(E_t(s_t)) + \beta \mathbb{E}[U_{t+1}]
\end{equation}

\noindent Where:
\begin{itemize}
    \item $U_t$ denotes the total recursive utility at time \( t \), representing the cumulative value of aligned innovation across time.
    \item $E_t(s_t)$ is the \textit{experiential realisation function}, mapping the state of the system \( s_t \) into a vector of need-satisfying outcomes at time \( t \). This encapsulates the actualised contributions of science to human well-being, across all domains of experiential value, such as health, meaning, sustainability, peace, self-actualisation, and beyond.
    \item $u(\cdot)$ is the \textit{instantaneous utility function}, capturing the present-period utility derived from the realisation of experiential outcomes.
    \item $\beta \in (0,1)$ is the standard \textit{intertemporal discount factor}, reflecting the weight assigned to future experiential alignment relative to the present.
    \item $\mathbb{E}[U_{t+1}]$ is the expected utility of future experiential realisations, introducing dynamic optimisation under uncertainty.
\end{itemize}

\vspace{1em}
\noindent This formulation interprets science not as a neutral or purely instrumental process, but as a recursive alignment mechanism. The goal of scientific inquiry becomes the continuous refinement of states \( s_t \) such that their downstream experiential outputs \( E_t(s_t) \) better satisfy the full range of human needs, both material and existential. 

\noindent In this view, innovation is not merely the expansion of a cognitive frontier, but the dynamic actualisation of human flourishing over time. As ideation costs collapse and epistemic uncertainty transitions to tractable risk, the recursive alignment of AI-enhanced discovery with the experiential matrix becomes both the challenge and the opportunity of post-scientific progress.

\vspace{1em}
\noindent EMT thus offers a definitive answer to the foundational question, \textit{What is science ultimately for?} In the post-science paradigm, the answer is clear. Science exists to align knowledge production with the lived, evolving, and multidimensional experience of being human. This recursive function encapsulates that normative purpose within a tractable formal structure.

Institutional implications derive from this. Systems must be built to update recursively. Funding models, publication structures, and researcher incentives must shift from static project proposals to flexible, dynamically re-optimising architectures that reward alignment, adaptability, and continuous responsiveness. \textit{EMT formalises a link between growth and ethics, in that it reconceptualises growth as the expansion of a dynamic matrix of human experiential needs}. This might be considered an \textit{ethical anchor} in growth theory terms, and a \textit{ rational for alignment between AI and human values and goals}, the common objective being the ever-improving human experiential need satisfaction. 

\section{Post Science Ethics, AI Alignment, and Ideation Stack Governance}

A collapse in ideation costs, if unmanaged, could overwhelm society with misaligned or low-value knowledge. As AI agents increasingly participate in ideation and discovery, governance of the ``ideation stack”, the layered architecture of idea generation, selection, validation, and application, becomes essential. We argue that alignment must be defined experientially, governed institutionally, and enacted recursively. These recommendations are of course offered as exploratory proposals rather than prescriptive solutions. They aim to prompt critical reflection and guide future dialogue, recognising the inherent uncertainties of a rapidly evolving AI landscape.

\subsection{AI Safety as Experiential Alignment}

Traditional AI safety frameworks focus on avoiding catastrophic outcomes or logical misalignments. In contrast, EMT suggests that the highest standard of alignment is experiential and that ideation should be evaluated based on its capacity to satisfy evolving, dynamic human needs.

Let \( \mathcal{I}(t) \) be the ideation stream, and \( U_{EMT}(i) \) the utility function defined over idea \( i \). Alignment then requires:
\[
\forall i \in \mathcal{I}(t), \quad U_{EMT}(i) > \delta
\]
where \( \delta \) is a normative threshold for experiential value. This reframes AI safety from a logic problem to a moral-epistemic one, anchored not in fixed objectives but in continuous alignment with human flourishing.

Experiential alignment is therefore considered a moral-epistemic criterion for AI safety. As discussed, we define $I(t)$ as the ideation stream at time $t$, representing the continuous emergence of novel ideas generated by human intelligence and AI. Each idea $i \in I(t)$ is evaluated using a utility function $U_{EMT}(i)$, aligned with the principles of EMT, which assess the contribution of each idea to core dimensions of human flourishing, including health, sustainability, meaning, equity, and self-actualisation. 

The alignment condition $\forall i \in I(t), \, U_{EMT}(i) > \delta$ imposes a normative threshold $\delta$ for experiential value, ensuring that only ideas which exceed a minimum standard of positive impact are considered acceptable. This reframes AI safety as a moral-epistemic challenge rather than a purely logical constraint problem, whereby the objective is not to hard-code fixed goals, but to continuously align ideation outputs with evolving, pluralistic, and dynamic conceptions of human flourishing. It demands an ongoing evaluative process grounded in ethical judgement and epistemic sensitivity, acknowledging that the value of ideas must be assessed not in abstract, but in relation to their real-world experiential consequences. In this light, ideation becomes not only a technical process but a moral act, and AI safety becomes inseparable from the continuous alignment of innovation with the lived needs and values of humanity.

\medskip

\vspace{0.5em}

While classical AI safety emphasises logical consistency and the avoidance of unintended outcomes (e.g., the ``paperclip maximiser” problem), this formulation therefore draws attention to a deeper concern, the alignment of AI ideation with the subjective and expanding set of human needs, understood through the lens of experiential value. 

Accordingly, as discussed, the threshold \( \delta \) represents the minimum acceptable contribution to human wellbeing, interpreted broadly across domains such as emotional fulfillment, cognitive development, social connection, and existential meaning.

This shift in framing is significant. Rather than viewing AI alignment as a fixed or static problem that can be solved through hardcoded goals or constraints, EMT positions it as an ongoing process of evaluative alignment with an evolving experiential matrix. In this view, the ``safety” of an AI system is not defined merely by the absence of harm but by the presence of positive, meaningful contribution to the human condition. This necessitates a normative benchmark that is dynamic, ethically informed, and sensitive to human development across time.

The ideation stream \( \mathcal{I}(t) \), representing the real-time flow of ideas or actions generated by AI, therefore becomes a space of constant moral evaluation. Each idea is scored by its ability to satisfy not just functional needs (e.g., speed, cost reduction) but also higher-order needs including belonging, purpose, creativity, and self-actualisation. Through its ability to ingest and interpret the entirety of global output in real time, AI enables unprecedented evaluative capacity.  EMT thus reframes AI alignment as a recursive alignment problem, one in which ideation is guided by feedback loops between human experience and machine outputs.

This formulation is not only theoretically distinct but practically urgent. As AI systems increasingly participate in knowledge creation, recommendation engines, education, and health, the question is no longer only whether they behave as intended, but whether their outputs actively co-create a flourishing experiential world. Thus, AI safety becomes synonymous with human-aligned meaning-making. In this respect, EMT does not discard classical safety logic but subsumes it within a broader framework of normative human alignment.

\subsection{Governance of Ideation Stacks}

As discussed, we define an \textit{ideation stack} as a layered architecture of AI agents, human experts, feedback systems, and interpretability mechanisms that collectively generate and evaluate new knowledge. The ideation stack spans multiple layers, such as data ingestion, model formation, hypothesis generation, prioritisation, experimentation, and dissemination. The stack includes base models trained on scientific corpora and experiential data; agentic researchers capable of autonomous hypothesis generation; human alignment layers for steering, correction, and sensemaking; and feedback protocols from stakeholders, policymakers, and affected communities.

Each layer of the ideation stack introduces new alignment risks, but also new governance opportunities. In low-stakes domains, lightweight oversight may suffice. In high-stakes or irreversible domains (e.g., biomedical interventions, artifical general intelligence (AGI) development), full-stack transparency and participatory deliberation become essential.

Each layer also introduces the potential for epistemic drift or ethical misalignment. Therefore, governance must be multi-tiered, incorporating \textit{upstream oversight} to ensure training data and prompts are ethically sourced and bias-minimised, \textit{midstream filters} to facilitate real-time curation mechanisms that prioritise needs-aligned ideation, and \textit{downstream validation} to provide human-in-the-loop evaluation that includes diverse perspectives and experiential metrics.

This stack must operate under principles akin to algorithmic transparency, peer accountability, and recursive auditability.

\subsection{Institutional Mechanisms for Human-AI Epistemic Partnership}

As ideation becomes increasingly AI-mediated, the role of human researchers must evolve from that of sole originators of knowledge to that of curators, stewards, and ethical filters within an expanded epistemic system. This shift requires the development of novel institutional mechanisms capable of ensuring that human agency remains central to the evaluation and direction of AI-generated ideation, especially in the context of EMT, which foregrounds the normative importance of aligning knowledge with dynamic human experiential needs.

Building upon this, we propose the creation of \textit{EMT alignment panels}, in the form of interdisciplinary, continuously adaptive bodies tasked with assessing AI-generated ideation not through conventional bibliometric metrics such as citation counts, but through their alignment with evolving experiential matrices. These panels, comprising ethicists, domain experts, data scientists, and public representatives, would be responsible for curating and updating the experiential matrices \( E_t \) based on new empirical data, societal shifts, and evolving ethical priorities. In doing so, they anchor AI ideation models to a dynamic representation of human needs and values. This role is not static but iterative, allowing ideation to remain responsive to the plural, lived dimensions of human flourishing.

In parallel, we envision the development of \textit{dynamic oversight algorithms} designed to regulate and steer the ideation process in real time. These systems might incorporate feedback loops derived from human experience data, including emotional, affective, and relational indicators, thereby enabling ideation to be updated and filtered in response to changing social contexts. Such algorithms might also ensure that AI does not merely optimise static objectives, but aligns ideation with the broader human condition, operationalising the moral-epistemic orientation at the core of EMT.

To further institutionalise reflexivity, we advocate for the replacement of traditional static peer review structures with \textit{recursive review protocols}. These protocols would institute a multi-layered evaluative architecture in which all knowledge outputs undergo recursive, multi-agent critique. This would include all possible human and AI interrogation, and participatory feedback loops involving diverse stakeholders to anchor this to alignment with human experiential needs. This recursive structure mirrors the theoretical foundations of EMT, where knowledge is seen as an evolving alignment game between ideation and lived experience, requiring ongoing calibration rather than one-time validation.

Transparency is further supported through the implementation of \textit{public audit layers}. All ideation outputs exceeding a defined threshold of social, ethical, or epistemic significance must pass through publicly visible audit trails. These trails document key epistemic choices, model assumptions, and the rationale for alignment judgments. Such logs allow ideation to be made contestable and revisable, facilitating democratic oversight and ensuring that ideation processes remain accountable to a plurality of worldviews. This mechanism guards against monocultural drift, where optimisation systems converge on narrow or homogenised metrics of value.

Taken together, these institutional proposals reposition ideation as a reflexive and democratically governed process. They operationalise EMT's core claim that ideation must not only be productive in terms of output generation but also reflective, and sensitive to the heterogeneous values, moral horizons, and experiential needs of diverse human populations. These mechanisms are designed not simply to improve AI systems, but to embed AI ideation within an epistemic ecology structured by recursive governance, ethical deliberation, and participatory norm-setting.

In this context, human agency does not recede but becomes more central than ever. It is indispensable not only for technical evaluation and strategic oversight but also for protecting the normative integrity of knowledge. Human actors serve as the epistemic conscience of the system, ensuring that the accelerating ideation capacity enabled by AI does not outpace its moral intelligibility or societal legitimacy. As such, the human–AI partnership must be institutionalised not merely as a technical arrangement but as a shared epistemic enterprise rooted in ethical responsibility and collective flourishing.

\subsection{EMT-Aligned Oversight Architecture}

An oversight architecture grounded in EMT reorients governance away from a narrow focus on formal correctness or technical efficiency, and instead centres on what we term \textit{experience-sensitivity}, the continuous evaluation of ideation outputs in terms of their anticipated impacts on the evolving spectrum of human experience. Validity, reliability, and trustworthiness, once central concerns in the traditional scientific paradigm, are seamlessly and near-comprehensively achieved within the post science regime, where AI-driven evaluation, real-time feedback, and system-level alignment converge to ensure consistently rigorous outputs.

In this paradigm, the ultimate criterion for implementation is not only whether an output is logically sound or economically viable, but whether it contributes meaningfully to human flourishing in diverse and situated ways.

Let \( \Phi \) denote the full ideation stack, the set of all candidate ideas, models, designs, or policies generated by human or AI agents within a given decision cycle. Let \( O_t \subseteq \Phi \) represent the ideation outputs available at time \( t \). These outputs may be scientific claims, technological artifacts, or governance proposals. Let \( E_t \) denote the experiential matrix at time \( t \), a dynamic structure representing the collective state of human needs, values, capabilities, and lived conditions, derived from empirical indicators and normative deliberation. Finally, let \( \alpha(t) \) represent the societal alignment vector at time \( t \), capturing ethically weighted priorities such as sustainability, justice, autonomy, and well-being.

We define EMT-aligned governance as a filtering function:
\begin{equation}
\mathcal{G}_\text{EMT}(O_t, E_t, \alpha(t)) \rightarrow O^*_t,
\end{equation}
where \( \mathcal{G}_\text{EMT}(\cdot) \) is a value-sensitive governance mechanism that maps the full set of ideation outputs \( O_t \) into a selected subset \( O^*_t \subseteq O_t \) deemed suitable for implementation, further development, or social deployment. This function filters outputs not only based on feasibility or cost-efficiency, but on their projected contribution to the multidimensional space of human flourishing as encoded in \( E_t \) and steered by \( \alpha(t) \).

To operationalise this architecture, we propose a three-layered structure that instantiates EMT’s normative commitments.

First, a \textit{normative kernel} defines the core evaluative principles by which outputs are assessed. This kernel includes criteria such as well-being, equity, sustainability, and dignity, each understood not as abstract ideals but as context-sensitive vectors encoded within the experiential matrix. This ensures that ethical reasoning is structurally embedded within the governance function, rather than externally appended.

Second, a \textit{recursive monitoring layer} enables continuous, real-time feedback by drawing on traceability logs from AI systems, user interactions, and participatory community responses. This layer allows the experiential matrix \( E_t \) to be updated dynamically, such that governance decisions remain responsive to lived experience rather than fixed metrics. Recursive monitoring helps prevent drift toward optimisation proxies that may unintentionally neglect or harm certain populations.

Third, a \textit{deliberative adaptation layer} provides institutional grounding through regular review by interdisciplinary panels composed of ethicists, economists, technologists, and members of the public. These panels are responsible for refining the alignment vector \( \alpha(t) \) through structured deliberation, democratic participation, and horizon scanning. In doing so, they maintain normative relevance across social and temporal contexts.

Taken together, this EMT-aligned oversight architecture transforms governance into an epistemic and ethical filtering process, one that \textit{recognises ideation as a form of worldmaking}. Rather than assuming that technically feasible outputs are inherently desirable, it asserts that outputs must be judged in relation to their capacity to expand, deepen, or safeguard the space of human possibility. This approach enacts the core principle of EMT, that economic and epistemic systems must co-evolve in alignment with the advancing frontier of human experiential needs.

For both expert and non-expert readers, the implication is straightforward but profound. We should not only ask whether an AI-generated idea is smart or correct (this in a post science world is a given), but whether it makes the world more livable, meaningful, and just. EMT provides a principled framework to embed this question at the heart of oversight design.

This system provides a scalable blueprint for post science knowledge governance, where AI does not merely produce ideas, but participates in an ethically governed ecosystem of ideation aligned with experiential value.

In sum, AI alignment in the post science era cannot be outsourced to technical fixes alone. It requires a moral architecture of ideation governance, one that integrates recursive optimisation, experiential utility, and participatory oversight. This is not a constraint on science; rather, it is its ethical completion.

\medskip

\noindent These sections have highlighted the importance of recognising that as ideation costs approach zero, the bottleneck in science and innovation shifts from epistemic capacity to ethical alignment. They have stressed that in a post-science paradigm, where AI systems continuously generate hypotheses, test designs, and policy suggestions, the crucial question becomes not \textit{what} can be known, but \textit{what should be pursued}. In doing so, they have framed AI safety and AI ethics as a problem of experiential alignment, introducing the concept of the ideation stack, and outlining governance mechanisms that might be helpful in ensuring that AI–human partnerships remain epistemically robust and ethically grounded.

Ultimately, science in the post-science paradigm therefore becomes not merely a knowledge-producing enterprise, but a goal-oriented, ethically governed, dynamically updating partnership between human and machine intelligence. The concluding sections now turn to implications of the potential for these components to cohere into a full innovation system, one capable of continuous ideation under conditions of uncertainty collapse and experiential alignment.

\section{The Epistemic Inversion of the University: Toward a Post-Scarcity Knowledge Institution}

The university, as historically constituted, emerged under conditions of profound knowledge scarcity. Its structure, rituals, and credentialing systems were built to preserve, curate, and transmit knowledge in a world where such resources were limited, tacit, and slow to diffuse. But the epistemic landscape has inverted. AI now allows for real-time comprehension, synthesis, and generation of knowledge at scales that dwarf traditional systems. In this post science paradigm, knowledge is no longer scarce. What is scarce is alignment.

The university, like the journal system, was designed for a world where knowledge was scarce and its acquisition slow, costly, and labor-intensive. Its legitimacy rested on epistemic privilege, the ability to access, transmit, and certify rare knowledge. But with the collapse in ideation costs, AI has inverted this premise. The knowledge-value axis of civilization is being fundamentally reversed.

Modern universities emerged to meet the needs of a world in which knowledge was scarce, tacit, and difficult to generate. Their structures, of tenure, faculties, academic publishing, and credentialing, reflect the historical assumption that truth is hard-won and should be slowly accumulated under institutional guardianship. 

AI, by collapsing the cost of ideation, fundamentally inverts this previous logic. In the post science paradigm, the foundational value proposition of the university as a knowledge custodian no longer holds.

This is not merely a technological shift, it is an epistemological inversion. The \textit{scarcity of knowledge} has been replaced by the \textit{scarcity of alignment}. What is now rare is not information, but judgment. Not data, but discernment. The core principle of scientific progress, its ability to generate new knowledge, must now be reoriented toward directing that knowledge in ways that serve human flourishing.

\subsection{From Knowledge Scarcity to Knowledge Saturation}

In the traditional paradigm, knowledge was rare, and credentialed institutions were necessary to organise and validate it. Now, AI agents can parse, synthesise, and generate new knowledge at scales and speeds that defy human capacity. Knowledge is no longer scarce, but knowledge saturation prevails. This inverts the foundational assumption of universities, that knowledge accumulation is intrinsically difficult and slow.

We call this the \textit{paradox of epistemic inversion}, whereby that which was once rare (insight) becomes abundant, and that which was once undervalued (judgment, alignment, care) becomes scarce and precious.

\subsection{Reimagining the Role of the University}

In the post science paradigm, the value of the university lies not in transmitting content, but in facilitating context. We propose new function for the university, as alignment institutions, where students learn to interpret AI-generated knowledge ethically, systemically, and experientially; as human-AI laboratories, where teams of humans and AI systems co-produce knowledge in ways aligned with community needs; and as social feedback engines, where lived experience refines the direction of inquiry, reorienting knowledge production around dynamic need frontiers.

Rather than centers of scarcity, universities can become hubs of ethical coordination in the abundance economy. Their legitimacy would no longer rest on exclusive access, but on inclusive alignment.

This would require post-scarcity educational design. To remain relevant, universities must invert their own logic. This implies shifting from curriculum-based to need-responsive learning architectures, embedding AI fluency, ethical reasoning, and experiential modeling at the core of every program, and valuing care roles, guidance, and epistemic stewardship at the top of academic labor hierarchies.

The greatest contribution a university can now make is not to produce knowledge, but to humanise, align, and guide it. In the post science paradigm, that is the frontier of educational value.

Education, long premised on the scarcity of knowledge, must now therefore be redefined in the context of epistemological inversion. In the traditional model, the university served as the gatekeeper of information, authority flowed from content expertise, and pedagogy revolved around the transmission of scarce facts and methods. 

However, in a post science paradigm enabled by AI, knowledge becomes abundant, instantly generatable, and ubiquitously accessible. The fundamental scarcity shifts, not from information, but to alignment. 

The purpose of education therefore undergoes a radical transformation, from the curation of content to the cultivation of alignment. Cognitive alignment becomes essential, as students must now discern signal from noise in AI-generated ideation. 

Ethical alignment becomes central, ensuring that outputs and innovations serve human flourishing rather than reinforce harm or misalignment. 

Experiential alignment, the heart of EMT, emerges as the highest educational aim, connecting ideation with the evolving, infinite matrix of human needs. Educators must evolve into alignment mentors, needs interpreters, and curators of purpose. \textit{This reframing rescues education from obsolescence}. Rather than being displaced by AI, education becomes the very site where AI is aligned, where meaning is cultivated, and where humanity learns to grow with integrity. 

In this new paradigm, the university becomes a compass, not a vault. It teaches not what to think, but how to align thinking itself with a future worth living. 

Thus, education retains a vital role. It is no longer about transmitting knowledge, but \textit{about training alignment with the truths and needs that matter most in a post-scarcity ideational landscape}.

The greatest contribution a university can now make, therefore, is not to produce knowledge, but to \textit{humanise, align, and guide it}. In the post science paradigm, that is the true frontier of educational value.

In parallel, the broader economy must also undergo a structural inversion. In a post-scarcity world, value no longer flows from the control of scarce goods but from alignment with an ever-expanding frontier of human experiential needs. State subsidies must play a catalytic role in this transition. \textit{Rather than reinforcing outdated hierarchies of labour value based on scarcity or market pricing, public policy might consider}:

\begin{itemize}
    \item Ensuring that care workers, those who maintain the human condition and emotional infrastructure of society, are among the most valued and highest-paid.
    \item Incentivising contributions to the highest-order human needs (e.g., meaning, connection, purpose, transcendence) through direct support and recognition.
    \item Rewarding roles that enable alignment, empathy, and the cultivation of flourishing over those that simply extend production.
\end{itemize}

Such mechanisms would gradually reshape the structure of jobs, wages, and status to reflect the logic of EMT. The result is the emergence of an \textit{alignment economy}, a system in which AI, human labor, and capital converge to fulfil the deep matrix of human needs rather than merely driving GDP growth.

Universities in this economy become coordination hubs, not just of knowledge, but of moral vision, experiential design, and human-aligned futures. They no longer merely transmit what is known but participate actively in shaping the possible.

Accordingly, the greatest contribution a university can now make is not to produce knowledge, but to humanise, align, and guide it. In the post science paradigm, that is the frontier of educational value.

\subsection{Post-Scarcity Educational Design}

With AI capable of ingesting, synthesising, and recombining all available literature in real time, the constraint has moved upstream from knowledge creation to problem selection and downstream to need-aligned application. In this environment, knowledge has become commoditised, and no longer a differenting factor between individuals or institutions.

Universities can no longer compete on the basis of information control or authoritative dissemination alone. Their value lies in human development, in terms of ethical reasoning, deep listening, experiential interpretation, and social coordination.

As discussed, to remain relevant, universities must become platforms for aligning ideation with human experience. This includes prioritising curricula that build \textit{relational and ethical capacities}, including empathy, care, stewardship, and community building; becoming \textit{experience labs} where AI and humans co-design solutions to real-world needs in recursive feedback loops; focusing on \textit{alignment education}, training humans to discern what forms of knowledge creation are socially valuable; and evolving toward \textit{epistemic cooperatives} that integrate AI agents, researchers, practitioners, and citizens in need-responsive coalitions.

These are not tweaks to pedagogy, they are deep shifts in institutional purpose. The university must now become a custodian not of knowledge, but of alignment.

\subsection{From Prestige to Stewardship}

In a post-scarcity knowledge regime, prestige-based academic hierarchies may lose relevance. What replaces them are credibility economies anchored in \textit{demonstrated alignment with evolving human needs}, \textit{responsiveness to ideation opportunities as they emerge}, and \textit{real-time collaboration across institutional, geographic, and disciplinary boundaries}.

The most valued educational institutions may no longer be those with the most publications, but those with the strongest records of turning knowledge into flourishing.

\subsection{Epistemological Inversion as Civilizational Pivot}

This epistemological inversion may prove \textit{as consequential as the scientific revolution itself}. It challenges centuries of assumptions about what constitutes value, credibility, and authority. In its wake, institutions must transform, or risk becoming artifacts of a slower epistemic era. Universities must shift from being temples of accumulated knowledge to engines of real-time, needs-driven meaning-making.

If their core function remains defined as knowledge transfer or disciplinary gatekeeping, they risk obsolescence within a decade. Already, AI-powered tutors, ideation engines, and distributed learning platforms are outperforming average institutional offerings. Research, once a marker of institutional prestige, is increasingly undertaken in real time, by AI-human systems operating outside the conventional peer-reviewed journal pipeline. The prestige of credentials is eroding as firms and governments begin to recognise high-signal, AI-augmented learning portfolios.

But there remains a critical role for institutions, if they are willing to transform. In the post science era, as discussed, the university must pivot from being a repository of knowledge to a steward of alignment. Accordingly, its enduring value will lie in its ability to guide ethical alignment between human needs and machine capability; facilitate transdisciplinary coordination across problem domains; curate experiential learning environments that foreground wisdom, judgment, and care; and support the governance of real-time, AI-accelerated research ecosystems.

The university as a physical and symbolic space can survive, but only by transforming into an institution that adds value not through what it teaches, but through how it aligns. In the age of epistemic inversion, the most valuable faculty may no longer be disciplinary experts, but those capable of curating meaning, facilitating alignment, and anchoring exponential knowledge systems to human needs.

The countdown has begun. The university has, perhaps, at most a decade to reconstitute its function for the AI era. What survives will not be the institution in its current form, but the moral and cognitive scaffolding it offers, reimagined as an engine for alignment-centered worldmaking.

\subsection{Incentivising the Alignment Economy: Valuing Human Needs in a Post-Scarcity Paradigm}

As repeatedly argued throughout this work, as ideation costs collapse under AI, the primary constraint on innovation is no longer scarcity of knowledge, but the alignment of output with dynamic, multidimensional human needs. In this section, we formalise the role of the state in orchestrating a transition from a scarcity-based economy to what we envision as an \textit{alignment economy}, where occupations are valued based on their contribution to human flourishing, rather than their scarcity in the market.

\subsubsection{Economic Rationale for Subsidy-Based Realignment}

Under EMT, societal welfare is maximised not by GDP alone, but by the satisfaction of an expanding frontier of human needs, both material and experiential. However, in market economies, care workers and those who guide or steward human development, \textit{roles which contribute profoundly to the matrix of human wellbeing}, are systematically undervalued due to low market-clearing wages. To correct this, we propose a state-driven reallocation of labor incentives using targeted subsidies. We suggest the following model of alignment of labor with experiential value. 

Let there be $n$ occupational categories indexed by $i = 1, ..., n$, with each occupation supplying $L_i$ units of labor. The government has a fixed subsidy budget $B$, which it uses to subsidize labor in different occupations. Let:

\begin{itemize}
    \item $w_i$ be the base wage (market wage) for occupation $i$
    \item $s_i$ be the per-unit labor subsidy provided by the state to occupation $i$
    \item $\Lambda_i$ be the alignment coefficient for occupation $i$, measuring its contribution to human experiential value
    \item $L_i(s_i)$ be the labor supply to occupation $i$, increasing in $s_i$
\end{itemize}

We assume workers reallocate across occupations in response to total compensation ($w_i + s_i$), and that their aggregate utility depends on how well the labor structure aligns with the experiential needs of society.

Accordingly, we define a social planner’s problem as maximizing total experiential alignment subject to a fiscal budget constraint on subsidies:

\begin{equation}
\max_{\{s_i\}} \sum_{i=1}^{n} \Lambda_i \cdot L_i(s_i)
\end{equation}
subject to
\begin{equation}
\sum_{i=1}^{n} s_i \cdot L_i(s_i) \leq B
\end{equation}

This formulation models the optimal allocation of subsidy support across occupations to steer labor toward roles with high alignment value $\Lambda_i$.

We now derive functional form assumptions for labour supply. Accordingly, we assume labor supply responds log-linearly to total compensation:

\begin{equation}
L_i(s_i) = \bar{L}_i \cdot \left(1 + \eta_i \cdot \ln(1 + s_i/w_i)\right)
\end{equation}

where $\bar{L}_i$ is the baseline labor supply to occupation $i$ without subsidy, and $\eta_i$ is the elasticity of labor supply with respect to effective wage.

Substituting the labor supply function into the objective and constraint, the planner’s problem becomes:

\begin{align}
\max_{\{s_i\}} & \sum_{i=1}^{n} \Lambda_i \cdot \bar{L}_i \cdot \left(1 + \eta_i \cdot \ln(1 + s_i/w_i) \right) \\
\text{s.t.} \quad & \sum_{i=1}^{n} s_i \cdot \bar{L}_i \cdot \left(1 + \eta_i \cdot \ln(1 + s_i/w_i) \right) \leq B
\end{align}

This is a constrained nonlinear optimisation problem. The first-order conditions yield optimal subsidy allocations $s_i^*$ that balance three forces:
\begin{itemize}
    \item The alignment value $\Lambda_i$ of the occupation.
    \item The responsiveness $\eta_i$ of labor supply to subsidies.
    \item The efficiency of subsidy use under budget constraint $B$.
\end{itemize}

Intuitively, high subsidies should be directed toward occupations with high $\Lambda_i$ (e.g., care work, experiential guidance, higher-order need satisfaction), and whose labor supply is responsive to such support.

Certain policy implications derive from this. 

Over time, if the subsidy vector $\{s_i^*\}$ is continuously updated based on empirical estimates of $\Lambda_i$ and $\eta_i$, the occupational structure of the economy gradually tilts toward alignment. This achieves an \textit{ethical inversion} of the labor market, in which historically undervalued roles become central to economic and social value; a redefinition of ``growth” toward satisfaction of needs, rather than production volume, and a post-scarcity economic regime in which the state acts as an alignment mechanism.

Under EMT, the purpose of economic systems is redefined. Growth is not measured by GDP alone, but by the progressive satisfaction of an expanding frontier of human needs, encompassing both material and higher-order experiential dimensions such as care, creativity, meaning, and self-actualisation. As discussed, in market economies, however, occupational roles that contribute most directly to experiential flourishing, such as caregiving, education, and psychosocial development, are often systematically undervalued due to low market-clearing wages. 

EMT proposes a governance mechanism in which the state can act as an alignment device, reallocating labor incentives via targeted subsidies. As discussed formally above, defining an alignment coefficient $\Lambda_i$ for each occupation $i$, representing its contribution to the experiential matrix, and modeling labor supply responses to effective compensation, the framework formalises a constrained optimisation problem, maximising total alignment $\sum_i \Lambda_i L_i(s_i)$ subject to a fixed subsidy budget $B$. This approach enables a principled, empirically grounded realignment of labor structures toward ethically weighted human priorities. Over time, continuous updating of alignment coefficients $\{\Lambda_i\}$ and responsiveness parameters $\{\eta_i\}$ allows for an adaptive reconfiguration of the occupational landscape. This yields an ethical inversion of the labor market in which undervalued but high-impact roles become central, and, as discussed, a redefinition of``growth” in terms of experiential value creation rather than output volume, supporting a post-scarcity economy anchored in human flourishing.

\subsubsection{Application to the Role of the University}

Universities, under this regime, serve as centers for training individuals in high-$\Lambda_i$ domains, such as care, ethical reasoning, stewardship, and AI-aligned epistemics; hosting dynamic modeling platforms to estimate and update $\Lambda_i$ empirically via surveys, AI inference, and wellbeing metrics; and becoming institutions not of knowledge scarcity, but of value alignment and human need synthesis.

Thus, the university and the state co-evolve into an ethical infrastructure for the alignment economy. The transition is not merely economic, but civilizational.

\section{Policy Implications: Governing the Alignment Economy}

The analyses in this paper suggest certain policy implications. The collapse of ideation costs triggered by advanced AI systems upends the foundational economic assumption of knowledge scarcity. As a result, a shift from a scarcity-based to an \textit{alignment-based} economic architecture is both necessary and urgent. This transition demands a fundamental rethinking of economic policy, labor structures, institutional roles, and incentive systems. Our conclusions and policy suggestions here fully align with the argument based on EMT principles that a universal income grant should immediately be instituted to support individuals affected by technological unemployment during this transition (Callaghan, 2025a; Callaghan, 2025b).


State subsidies might need to be reoriented to correct market failures in the valuation of labor that contributes most to experiential alignment. Occupations such as care work, education, epistemic guidance, and higher-order human development roles tend to exhibit high alignment coefficients ($\Lambda_i$) but may be structurally undercompensated. 
This mirrors longstanding findings in the economics of wellbeing, which emphasise that markets undervalue relational and affective labor. A national fiscal strategy should therefore prioritise these roles through targeted subsidies. AI systems can support this process by dynamically estimating $\Lambda_i$ values using real-time wellbeing data, feedback metrics, and social outcome tracking. Subsidies might then be optimally allocated via constrained maximisation models, as formalised in this paper.



Redesigning labor and wage structures constitutes a second core policy imperative within the alignment economy. Under EMT, the prevailing hierarchy, where capital-intensive, output-maximising roles are disproportionately rewarded over relational, affective, and meaning-producing work, must be fundamentally reversed. In a post-scarcity regime, roles tied to routine production are more susceptible to automation and obsolescence, while human-centered, care-oriented, and purpose-driven labor becomes increasingly essential. EMT thus prioritises caring work, for example, by definition, viewing it as central to the satisfaction of higher-order experiential needs and the long-term flourishing of individuals and communities. This calls for a deliberate restructuring of economic incentives to reflect the true experiential value of work.

Compensation should therefore reflect alignment with the experiential matrix rather than marginal productivity in physical output. Governments can facilitate transitions into high-alignment roles by funding retraining programs, universal service incentives, and AI-augmented onboarding pathways.
Broader economic performance should be measured using multidimensional wellbeing metrics and alignment indices, displacing GDP as the central indicator.

A third priority concerns the transformation of universities. As ideation becomes abundant and increasingly AI-automated, universities must transition from their historical function as knowledge transmission institutions to platforms for alignment coordination and epistemic stewardship.
Funding models should reflect this shift, rewarding contributions to social alignment and systems integration rather than publication volume or market-oriented performance indicators 
Universities should be encouraged to develop real-time experiential modeling systems that map societal needs, identify misalignments, and support collective prioritisation. 
AI fluency, ethical reasoning, and systems thinking should be embedded across all disciplines, not as peripheral literacies, but as foundational to higher education in the post-science paradigm. 


A fourth domain of policy innovation lies in AI-guided public foresight and innovation governance. Governments might develop AI-powered platforms that simulate long-term societal impacts of emerging technologies, enabling anticipatory and participatory decision-making. 
These systems should support open foresight processes in which citizens, experts, and AI collaboratively construct and evaluate plausible futures. 

Institutions might also be mandated to conduct regular epistemic risk audits, particularly for ideation-accelerating or alignment-sensitive technologies. 

Fiscal logic itself might also be reconceptualised. Taxation policy should gradually shift from income-based to externality-based regimes, penalising misaligned innovation and rewarding contributions to alignment. 


Public research and development expenditure should not only target the technological frontier, but also the alignment frontier, in the form of investments in institutions, tools, and theories that enable better satisfaction of complex and evolving human needs.


A National Alignment Fund may serve as a dynamic vehicle for such reallocation, updated continuously through AI-driven need-sensing systems and public consultation platforms. 


Finally, achieving coherence across this policy architecture requires the adoption of post-GDP metrics and, potentially, constitutional economic reform. Alignment should be formally enshrined as a national economic objective, on par with productivity or employment. Institutional mandates, such as those of central banks or regulatory agencies, might be updated to reflect the goals of epistemic resilience and human flourishing.


Alignment-weighted indices of economic output, subjective wellbeing, and systemic coherence should replace GDP as the principal metric guiding public policy.  


In sum, governing the alignment economy requires more than correcting for market imperfections. It demands the deliberate reconstruction of incentive structures, institutional missions, and economic logic itself. In this post-scarcity, post science era, economics must evolve from a descriptive discipline into a normative design framework, one that aligns technological capacity with the unfolding frontier of human needs to ensure the transition to the alignment economy.

\section{Conclusion: Toward an Economics of Alignment and the Alignment Economy in the Post-Science Paradigm}

This paper has presented a comprehensive reconceptualisation of growth, knowledge, and institutional function in an era defined by the collapse of ideation costs. We have argued that the traditional scarcity-based assumptions underpinning science, economics, and governance are undergoing a profound epistemological inversion. As AI systems reduce the cost and tacitness of ideation, the limiting factor in innovation shifts from knowledge creation to alignment with complex and evolving human needs. 

To respond to this inversion, we have developed a formal structure, in the form of EMT, that centers recursive human needs as the proper objective of innovation, and positions ideation as a means to serve that objective. EMT reframes the role of science not as the production of knowledge \textit{per se}, but as the continuous alignment of innovation with a multidimensional and expanding matrix of lived human experience. Within this framework, ideation becomes abundant, but alignment remains scarce. This reorients the normative logic of policy, requiring a shift from productivity maximisation to alignment optimisation.

We have demonstrated that if the theoretical assumptions of EMT hold, then this new logic necessitates the transformation of economic structures. A central implication is the need to redesign labor hierarchies and wage systems around experiential alignment rather than market productivity. Policies that subsidise high-alignment roles, particularly in care, education, and epistemic stewardship, become foundational to sustaining value in a post-scarcity economy. Universities, likewise, must invert their own epistemic logic, transitioning from knowledge gatekeepers to alignment infrastructures tasked with modeling societal needs, guiding innovation trajectories, and maintaining ethical and epistemic resilience in real time.

Our formal modeling also suggests that in such a world, the marginal value of ideation asymptotically approaches zero unless tethered to societal alignment. This provides not only a theoretical justification for policy intervention but a macroeconomic rationale for reconfiguring institutions around recursive feedback loops of human need. The model further implies that economic growth can remain unbounded if, and only if, the directionality of that growth tracks with the expansion of the human experiential matrix.

This post science paradigm calls for a new political economy. One in which AI is not merely an object of governance but a participant in guiding collective futures; one in which universities no longer act as slow-moving validators of knowledge, but as epistemic stewards of societal coherence; and one in which care, meaning, and complexity management are not residual economic functions but core drivers of productivity and wellbeing.

In a world where ideation is no longer scarce, the challenge is not how much we can know, but how well we can align. Economics, policy, and science must thus reconstitute their foundational assumptions. They must adopt a logic that is not only dynamic and recursive, but also deeply human in its orientation. This paper offers a first step in that reconstitution. It provides a defensible model, a policy framework, and an institutional vision for a future in which the alignment of innovation with experiential value is the primary organising principle of economic life.

\newpage
\section*{\LARGE References}

\hspace*{2em}Acemoglu, D. (2009). Introduction to modern economic growth, Princeton University Press, Princeton.

Aghion, P. and P. Howitt (1992). A Model of Growth Through Creative Destruction, Econometrica, 60, pp. 323-351.

Agrawal, A., J. McHale and A. Oettl (2018). Finding needles in haystacks: Artificial intelligence and recombinant growth. The economics of artificial intelligence: An agenda. pp. 149-174. University of Chicago Press.

Arrow, K. J. (1951). Social choice and individual values, John Wiley \& Sons, New York.

Arrow, K. J. and G. Debreu (1954). Existence of an equilibrium for a competitive economy, Econometrica, 22, pp. 265-290.

Callaghan, C.W. (2025a). A General Theory of Growth, Employment, and Technological Change: Experiential Matrix Theory and the Transition from GDP to Humanist Experiential Growth in the Age of Artificial Intelligence, arXiv:2505.19045.

Callaghan, C. W. (2015). Crowdsourced ‘R\&D’ and medical research, British Medical Bulletin, 115, pp. 67-76.

Callaghan, C. W. (2016). Disaster management, crowdsourced R\&D and probabilistic innovation theory: Toward real time disaster response capability, International Journal of Disaster Risk Reduction, 17, pp. 238-250.

Callaghan, C. W. (2018). Developing the transdisciplinary aging research agenda: new developments in big data, Current Aging Science, 11, pp. 33-44.

Callaghan, C. W. (2025b). Rethinking Growth and the Future of Work: The Alignment Economy as a General Theory of Human-Centered AI. SSRN 5269022. Cambridge.

Farmer, J. D. (2024). Making Sense of Chaos: A Better Economics for a Better World, Yale University Press, New Haven.

Farmer, J. D. and D. Foley (2009). The economy needs agent-based modelling, Nature, 460, pp. 685-686.

Forrester, J. (1961). Industrial Dynamics, MIT Press, Cambridge.

Grossman, G. M. and E. Helpman (1991). Quality Ladders in the Theory of Growth, The Review of Economic Studies, 58, pp. 43-61.

Hayek, F. A. (1945). The use of knowledge in society, The American Economic Review, 35, pp. 519-530.

Jones, C. I. (2023). Recipes and Economic Growth: A Combinatorial March Down an Exponential Tail, Journal of Political Economy, 131, pp. 1994-2031.

Kitchin, R. (2014). The Data Revolution: Big Data, Open Data, Data Infrastructures \& Their Consequences. London: SAGE Publications Ltd.

Knight, F. H. (1921). Risk, uncertainty and profit, Houghton Mifflin.

Kovalevskiy, O., J. Mateos-Garcia and K. Tunyasuvunakool (2024). AlphaFold two years on: Validation and impact, Proceedings of the National Academy of Sciences, 121, p. e2315002121.

Maslow, A. H. (1943). A theory of human motivation, Psychological Review, 50, pp. 370-396.

Maslow, A. H. (1970). Motivation and Personality, Harper and Row, New York.

Peirce, C. S. (1934). Collected papers of Charles Sanders Peirce, Harvard University Press, Cambridge.

Solow, R. M. (1956). A contribution to the theory of economic growth, The Quarterly Journal of Economics, 70, pp. 65-94.

Wiener, N. (1948). Cybernetics or Control and Communication in the Animal and the Machine, MIT Press, Cambridge.

\end{document}